\documentclass[authoryear,times,3p,12pt]{elsarticle}
\usepackage{hyperref}
\usepackage{chemformula}
\usepackage{graphicx,pdflscape,float}
\usepackage{subcaption}
\usepackage{wrapfig}
\usepackage{amsmath}
\usepackage{lipsum}
\usepackage{cleveref}
\usepackage{multirow}
\usepackage{multicol}
\usepackage{lscape}
\usepackage{cuted}
\usepackage{stfloats}
\usepackage{hyperref}
\usepackage{booktabs}
\usepackage{lineno} 
\usepackage{xcolor}
\usepackage{enumitem}
\usepackage{stackengine}

\journal{https://arxiv.org/}
\biboptions{sort&compress}

\begin{document}\sloppy

\begin{frontmatter}

  \title{{Simulations of near-source wind development and pollution dispersion over complex terrain under different thermal conditions}}

  \author[add1,add4]{Zhihao Li}
  \author[add1,add4]{Rebecca Tanzer-Gruener}
  \author[add1,add4]{Albert Presto}
  \author[add2,add3,add4]{Peter Adams}
  \author[add1,add4]{Satbir Singh\corref{mycorrespondingauthor}}
  \cortext[mycorrespondingauthor]{Corresponding author. 319 Scaife Hall, 5000 Forbes Avenue, Pittsburgh, PA 15213, United States.}
  \ead{satbirs@andrew.cmu.edu}

  \address[add1]{Department of Mechanical Engineering, Carnegie Mellon University, United States}
  \address[add2]{Department of Civil and Environmental Engineering, Carnegie Mellon University, United States}
  \address[add3]{Department of Engineering and Public Policy, Carnegie Mellon University, United States}
  \address[add4]{Center for Atmospheric Particle Studies, Carnegie Mellon University, United States}

  \begin{abstract}
    \nolinenumbers
    A computational fluid dynamics (CFD) model that solves the steady-state Reynolds-Averaged Navier-Stokes (RANS) equations using the $k-\varepsilon$ turbulence model for buoyant compressible pollution dispersion under different meteorological conditions is developed. A 6.4 $km$ by 6.4 $km$ computational domain over a complex terrain with a height of 1 $km$ above the ground surface is created. Meteorological data from multiple available sources are utilized to obtain boundary conditions of wind speed, air temperature, turbulent kinetic energy (TKE), and its dissipation rate. To evaluate the model, a monitoring network of four anemometers is deployed. Model predictions are compared with measurements of wind speed and the concentration of \ch{SO2} emitted by a local coke plant. Comparisons show that the predicted wind speeds are reasonably close to the measured mean wind speeds and the average error is within $10\%$ at one location where relative fast wind speeds are recorded. The CFD model also predicts the correct trend of varying wind speeds across multiple sites of different elevations. The model also provides good predictions of \ch{SO2} concentrations for multiple cases, considering the complex nature of the terrain and meteorological conditions.

  \end{abstract}

  \begin{keyword}
    Wind development \sep Pollution dispersion \sep Complex terrain \sep CFD simulation \sep Turbulence model
  \end{keyword}
\end{frontmatter}

%\linenumbers
\section{Introduction}

The atmospheric boundary layer (ABL) is the lowest part of the atmosphere whose behavior is directly influenced by Earth’s surface. Within the ABL, wind shear caused by the drag near the ground and vertical air movement as a result of buoyant forces generate turbulence. Turbulence modeling is vital for correct simulations of the momentum, heat, and mass transport within the ABL. Different CFD models have been widely applied to various engineering problems. For simulations of flow within the ABL, examples include the predictions of pollution dispersion \citep{Huser1997,Sladek2007,Pontiggia2009,Amorim2013,Bonifacio2014,Tominaga2018}, energy production of wind farms \citep{Dorenkamper2015,Dhunny2017}, the spread of wildfire \citep{Forthofer2014}, the assessment of pedestrian comfort in an urban environment \citep{Blocken2016}, etc.

However, most studies have only focused on flat terrain of uniform aerodynamic roughness length under neutral ABL. When it comes to real-world applications, there are more factors to consider. For example, simulations of ABL flow over complex terrain are necessary in many applications. A terrain is considered to be complex if it has irregular topography and variations in land use that will generate inhomogeneities in turbulence and winds. The thermal stability of the ABL can be divided into three classes: neutral, stable, and unstable. The neutral ABL only represents part of a typical diurnal cycle of ABL. Since buoyancy plays a key role under stable and unstable conditions, ignoring its effect may lead to errors in predictions of pollution dispersion. To be more specific, there are some common considerations that are needed when developing a CFD model to simulate near-source wind development and pollution dispersion within the ABL over complex terrain.

First, the inlet boundary conditions, such as mean wind speed and mean air temperature profiles as a function of vertical height, are usually generated according to the Monin–Obukhov similarity theory (MOST) with empirical parameters estimated from flat terrain \citep{Foken2006}. Such profiles are valid within the surface layer, which is only within 100 $m$ above the ground surface. However, these profiles are commonly applied up to 6000 $m$ in CFD simulations \citep{Breedt2018}. Besides, in most cases, the theoretical profiles are estimated from measurements at one height level \citep{Pieterse2013, Toparlar2019} using equations presented in \cite{Richards1993} with the horizontal homogeneity assumption (i.e. no gradients along stream-wise direction). These theoretical profiles may not be suitable for the inlet of complex terrains. \cite{Breedt2018} applied an artificial smoothing around the terrain  so that the inlet profiles can be applied on a completely flat terrain as it removes terrain features across boundaries. However, it is unclear if the different smoothing methods will lead to different predictions. \cite{Li2017} proposed two methods of determining the inlet boundary conditions: one is to fit the velocity into an empirical law, and the other one is to interpolate the values from a precursor simulation of the upstream region. The latter one required more computing power, and the improvement in accuracy was limited \citep{Li2017}.  Vertical wind and temperature profiles can be obtained from direct measurements (e.g. atmospheric sounding) and data assimilation (e.g. reanalysis products). Since the simulation results depend strongly on the boundary conditions, a methodology that utilizes all available data sources is needed to obtain good-quality vertical profiles of wind speed, temperature, and turbulence suitable for complex terrain. A methodology to obtain boundary conditions at the inlet will be proposed in the present work.

Second, turbulence models are used in CFD to predict the effects of turbulence on the flow field, and model constants or coefficients need to be specified. In most cases, atmospheric flow simulations are performed using the RANS equations together with the $k-\varepsilon$ turbulence model \citep{Launder1974}. The constants in the $k-\varepsilon$ model must be properly chosen in order to correctly simulate the effects of turbulence within the ABL \citep{Richards1993}.  Multiple sets of model constants for the $k-\varepsilon$ model have been used in the literature \citep{Launder1974,hagen1981simulation,crespo1985,Richards1993,Alinot2005,Bechmann2010,Richards2011,vanderLaan2016,Piroozmand2020}. For example, from experimental data and findings from \cite{crespo1985} under neutral ABL, \cite{Alinot2005} proposed the model constants of $C_{\mu}=0.033$ and $C_{\varepsilon 1}=1.176$ for the $k-\varepsilon$ model under all stability classes. \cite{Richards2011} used the standard value for $C_{\mu}=0.09$, but changed the von Kármán constant to 0.433 to satisfy horizontal homogeneity. There is less consensus over the specification of the model constant $C_{\varepsilon 3}$, which appears with the buoyancy team in the dissipation equation. Many different values of this constant ranging from -4.4 to 3.4 and many different equations to obtain its value have been reported in the literature for ABL flows \citep{crespo1985,Alinot2005,vanderLaan2016,Piroozmand2020}. Moreover, most studies of these model constants have been performed for flat terrain, and often under neutral conditions. In the present work, we will test the applicability of a particular set of $k-\varepsilon$ model constants for flow over a complex terrain.

Third, uniform aerodynamic roughness length is commonly used throughout the terrain surface \citep{Huser1997,Sladek2007,Forthofer2014}, even though the common aerodynamic roughness length may vary from 0.001 $m$ to 1.300 $m$ near pollution sources \citep{Cimorelli2005}. In CFD models, the effects of the actual roughness obstacles above the ground are generally included by using wall functions which are based on experiments with sand-grain roughness \citep{Schlichting2017}. For complex terrain, the aerodynamic roughness length of the obstacle is usually converted to an equivalent sand-grain roughness height by multiplying a factor of 30 \citep{Blocken2007}. This conversion implies that the half-height of the wall-adjacent cell needs to be at least the height of the equivalent sand-grain roughness. When generating the computational mesh for a region with a high aerodynamic roughness length (e.g. 1.0 $m$ for mountainous areas), the height of the wall-adjacent cell needs to be at least 60 $m$. The accuracy of CFD simulations can be compromised following such conversion \citep{Blocken2007}. \cite{Parente2011_2} proposed a new wall model based on aerodynamic roughness, which does not impose strict limitations in terms of near-wall grid resolution. In the present work, we will compare the predictions of the CFD model for uniform surface roughness and variable surface roughness for flow over a complex terrain using the aerodynamic-based wall model.

Fourth, after obtaining the steady-state flow field, we can start the simulation of a passive pollutant. Such a strategy has been applied in the studies of pollution dispersion over complex terrain that includes forest and urban areas \citep{Huser1997,Sldek2007,Gousseau2015, Karra2017, TojaSilva2018, Aristodemou2018}. In most cases, incompressible solvers with Boussinesq approximation are used and only the emission rate is specified without accounting for the effects of emission conditions. \cite{Amorim2013} used a incompressible CFD solver to model the dispersion of carbon monoxide emitted by road traffic under neutral condition and ignored the traffic produced turbulence. When simulating exhaust gas dispersion from chimneys of a power plant, \cite{TojaSilva2018} ignored the buoyancy effects of the exhaust temperature. It is well known that the Boussinesq approximation for ABL flow is relatively accurate as long as changes in the density are small \citep{Tominaga2018}. However, if the emission sources include hot stacks, the approximation may not be good. Ignoring the atmospheric stability or the local environment near the emission sources may result in totally different dispersion patterns.

Given the challenges of simulating atmospheric flow and pollution dispersion over complex terrain, the present work focuses on; 1) a method to construct a model for a complex terrain with a high-quality computational grid, 2) a method to generate reliable inlet boundary conditions with inputs from ground-level and vertically-distributed weather data sources, 3) model sensitivity to parameters such as terrain complexity, stack exit conditions, thermal stability class, and 4) investigation of sampling strategies to compare model-predicted pollutant dispersion with a single-point measurement.

\begin{figure}[hbt!]
  \centering
  %%trim=left bottom right top  
  \includegraphics[width=0.5\linewidth]{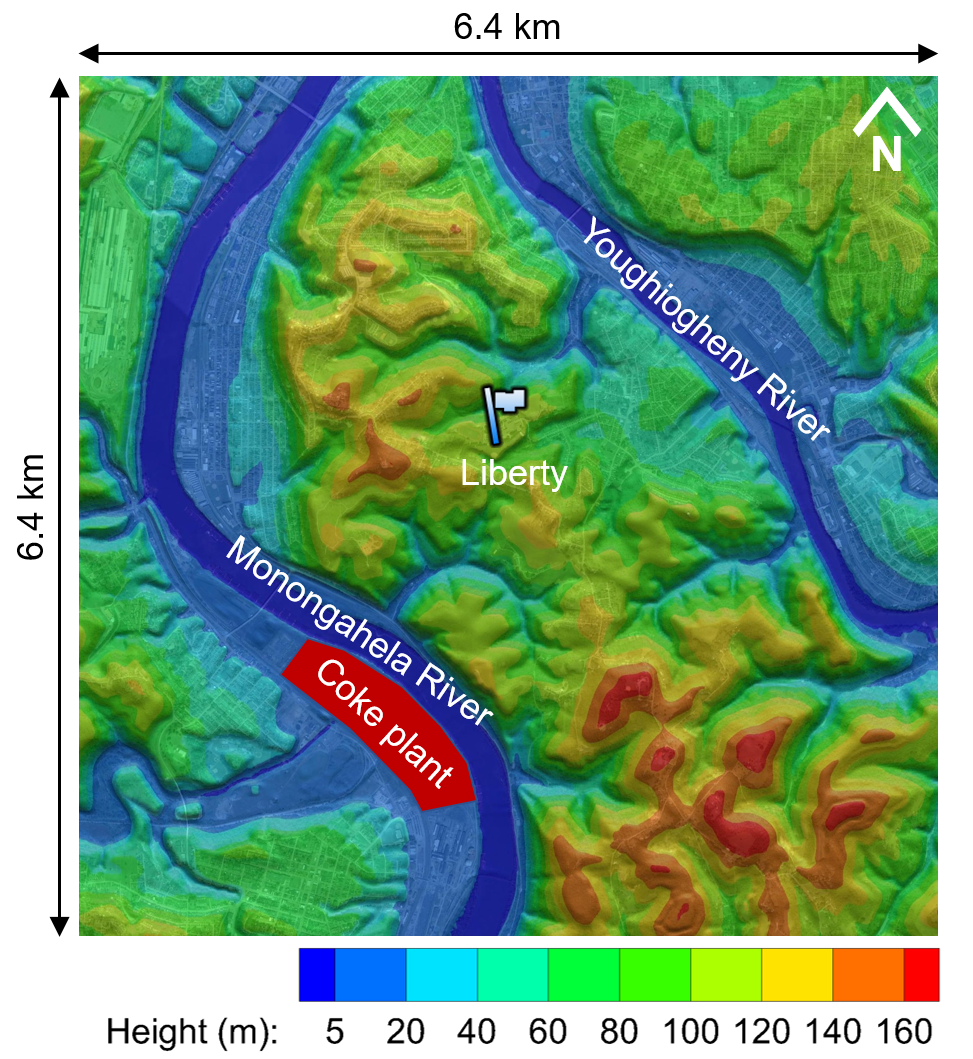}
  \caption{Height contours of the 6.4 $km$ by 6.4 $km$ study domain overlaid with satellite image from Google Earth. The domain is located approximately 15 $km$ south of Pittsburgh in southwestern Pennsylvania, U.S. }
  \label{fig:domain}
\end{figure}

The CFD model is applied for predictions of wind development and pollution dispersion over the Monongahela River valley region shown in Figure \ref{fig:domain}. There are two rivers flowing through this region: the Monongahela River is located on the west, and the Youghiogheny River is located on the east. The lowest point within the domain is located on the surface of the Monongahela River, and its height is set to 0 $m$. The highest point (174 $m$) is located on the mountain top near the southeast corner. The land-use types over this domain include industrial land, low-intensity residential area, and deciduous forest. Such irregular topography and land-use types make this terrain complex, compared to terrains previously reported in the literature. Predictions of wind development and pollution dispersion are of great interest in this study domain. The Clairton plant within the domain is the largest coke works in North America, and there is a history of high \ch{SO2} concentrations in the region, as recorded by the nearby Liberty monitoring station \citep{ACHD2019}. A CFD model that can provide accurate predictions of wind patterns and pollutant dispersion under different atmospheric stability classes will be useful to investigate the distribution of \ch{SO2} concentrations in the region.

\section{Model formulation}
\label{section:compressible}

The open-source CFD code OpenFOAM \citep{Weller1998} is used for the development of the CFD model. This section describes the equations related to the CFD model in the present work. Since the effects of buoyancy are important to this study, a steady-state solver for buoyant, turbulent flow of compressible fluids is chosen. Such a solver is already available in OpenFOAM and it is called ``buoyantSimpleFoam". By performing Reynolds averaging on the governing equations, the Reynolds-averaged Navier–Stokes equations (RANS) and other related equations can be obtained. The standard $k-\varepsilon$ model \citep{Jones1972} with additional buoyancy generation and dissipation terms is explained. The equation of state is used to relate pressure with density and temperature. Index notation is used in the equations within this section.

\subsection{Mass conservation}
For a compressible flow under steady state, the conservation law of mass gives the continuity equation:

\begin{equation}
  \frac{\partial }{\partial x_i} \left ( \rho u_i \right )= 0
\end{equation}
where $\rho$ is the density. $u$ is the mean velocity vector and $x$ is the Cartesian coordinate.
\subsection{Momentum equation}
The compressible steady-state momentum equation is:

\begin{equation}
  \frac{\partial}{\partial x_j} \left( \rho u_j u_i \right) =
  -\frac{\partial{p}} {\partial{x_i}} +\rho g_i  +  \frac{\partial}{\partial x_j} \left( \tau_{ij} + \tau_{t_{ij}} \right)
\end{equation}
where $p$ is the pressure. $g$ is the gravitational acceleration. $\tau $ is the mean stress tensor and $\tau_t $ is the Reynolds stress tensor. In this study, the focus is on the flow over a small region within only a few hundred meters above the ground, so the Coriolis force is neglected in the momentum equation. The pressure gradient and gravity force terms are rearranged in the following form:

\begin{align}
  \begin{aligned}
    -\frac{\partial{p}} {\partial{x_i}} +\rho g_i
     & = -\frac{\partial\left({p_\textup{rgh}} + \rho g_j x_j \right)} {\partial{x_i}} +\rho g_i                               \\
     & = -\frac{\partial\left( {p_\textup{rgh}}\right)}{\partial{x_i}}-\left(g_j x_j\right)\frac{\partial \rho}{\partial{x_i}}
  \end{aligned}
\end{align}
The new term $p_\textup{rgh}$ is called the pseudo hydrostatic pressure, which is the pressure in excess of the hydrostatic pressure over the total pressure $p$. $p_\textup{rgh}=p-\rho g z$, where $z$ is the height above the ground. This rearrangement is a common approach in CFD to simplify the pressure boundary conditions in the compressible flow solvers. Finally, the momentum equation becomes:

\begin{equation}
  \frac{\partial}{\partial x_j} \left( \rho u_j u_i \right) =
  -\frac{\partial\left( {p_\textup{rgh}}\right)}{\partial{x_i}}-\left(g_j x_j\right)\frac{\partial \rho}{\partial{x_i}}  +  \frac{\partial}{\partial x_j} \left( \tau_{ij} + \tau_{t_{ij}} \right)
  \label{eqn:momeq}
\end{equation}
\subsection{\texorpdfstring{$k$-$\varepsilon$}{k and e} turbulence model}
The $k-\varepsilon$ model is widely used to simulate turbulence in atmospheric flows. With the addition of the buoyancy effects, the model is called ``buoyantKEpsilon" in the OpenFOAM code. The transport equations for turbulent kinetic energy $k$ and its dissipation rate $\varepsilon$ are written as:

\begin{equation}
  \frac{\partial}{\partial x_i} (\rho u_i k) = \frac{\partial}{\partial x_j} \left[ \left(\mu + \frac{\mu_t}{\sigma_k} \right) \frac{\partial k}{\partial x_j}\right] + G_k + G_b - \rho \varepsilon
  \label{eqn:keqn}
\end{equation}
\begin{equation}
  \frac{\partial}{\partial x_i} (\rho u_i \varepsilon ) = \frac{\partial}{\partial x_j} \left[\left(\mu + \frac{\mu_t}{\sigma_{\varepsilon}} \right) \frac{\partial \varepsilon}{\partial x_j} \right] + C_{\varepsilon 1}\frac{\varepsilon}{k} \left( G_k + C_{\varepsilon 3} G_b \right) - C_{ \varepsilon 2} \rho \frac{\varepsilon^2}{k}
\end{equation}
$G_b$ is the production of $k$ due to buoyancy, which is given by:

\begin{equation}
  G_b=\beta g_i\frac{\mu_t}{Pr_t}\frac{\partial \theta}{\partial x_i}
  \label{gb}
\end{equation}
where $\beta$ is the coefficient of thermal expansion. $\mu_t$ is the kinematic viscosity. $Pr_t$ is the turbulent Prandtl number. $\theta$ is the potential temperature for air.
For stable stratification, buoyancy tends to suppress the turbulence resulting in $G_b<0$.
$G_k$ is the production of $k$ due to mechanical shear, which is given by:

\begin{equation}
  G_k= \tau_{t_{ij}} \frac{\partial u_i}{\partial x_j}
\end{equation}
The model constants proposed by \cite{crespo1985} and later adopted by \cite{Alinot2005} are used in this study: $C_{\mu}=0.033, C_{\varepsilon 1}=1.176$, and $C_{\varepsilon 2}=1.92$. As for the value of $C_{\varepsilon 3}$, there is less consensus in the literature. The constant $C_{\varepsilon 3}$ is multiplied with the buoyancy term, $G_b$. Different values of this constant have been used in the literature under different stability classes. Moreover, these values have been tested for atmospheric flows without the presence of hot plumes rising from stacks, as is the case in the present work. The hot plumes create a strongly buoyant local environment. For strongly buoyant flows, \citep{Henkes1991} proposed the following expression for $C_{\varepsilon 3}$:

\begin{equation}
  C_{\varepsilon 3}=\tanh \left |\frac{u_3}{\sqrt{u_1^2+u_2^2}}  \right |
\end{equation}
where $u_3$ is the vertical component of the flow velocity vector, and $u_1$ and $u_2$ are the horizontal components. We have adopted the above expression for $C_{\varepsilon 3}$ in the present work. This expression was also adopted by \citep{Piroozmand2020} for simulations of ABL flows.

\subsection{Equation for potential temperature}
\cite{Pontiggia2009} discussed that if the absolute temperature is employed, it is not possible to balance the adiabatic profile of absolute temperature by varying the pressure along the vertical direction. Therefore, it is convenient to solve for the potential temperature in the system of equations. The transport equation for the potential temperature is written as:

\begin{equation}
  \frac{\partial}{\partial x_i} (\rho u_i \theta) = \frac{\partial}{\partial x_i} \left( \rho \alpha_{\textup{eff}}\frac{\partial \theta}{\partial x_i}\right)
\end{equation}
where  the effective thermal diffusivity $\alpha_{\textup{eff}}$ is the sum of laminar diffusivity $\alpha$ and turbulent thermal diffusivity $\alpha_t$: $\alpha_{\textup{eff}} = \alpha + \alpha_t = \frac{\nu}{Pr} + \frac{\nu_t}{Pr_t}$. $\nu$ and $\nu_t$ are the kinematic viscosity and turbulent kinematic viscosity. In this study, the Prandtl number $Pr=0.7$ and turbulent Prandtl number $Pr_t=0.85$.

\subsection{Equation of state}
For perfect gas under the weak compressible assumption \citep{Huser1997}, the equation of state is:

\begin{equation}
  p = \rho R \theta
\end{equation}
where $R$ is the gas constant. With the equation of state, the production of $k$ due to buoyancy can be simplified to:

\begin{equation}
  G_b=-g_i\frac{\mu_t}{\rho Pr_t}\frac{\partial \rho}{\partial x_i}
\end{equation}

\subsection{Passive scalar transport equation}
Once emitted from their sources, pollutants quickly mix with the air. Unlike temperature or pressure, they do not have direct effects on the wind field in the atmosphere. If the chemical reactions are also ignored, they can be treated as passive scalars transported by the wind. To simulate the dispersion of a passive scalar $\phi$, which is \ch{SO2} in this study, the steady-state scalar transport equation is written as:
\begin{equation}
  \frac{\partial}{\partial x_i}\left(\rho u_i\phi\right)=\frac{\partial}{\partial x_i}\left(\rho D_\textup{{eff}}\frac{\partial\phi}{\partial x_i}\right)+S_\phi
  \label{eqn:scalar}
\end{equation}
where the effective mass diffusivity $D_\textup{{eff}}$ is the sum of laminar mass diffusivity $D$ and turbulent effective mass $D_t$: $D_\textup{{eff}}=D+D_t=\frac{\nu}{Sc} + \frac{\nu_t}{Sc_t}$. The molecular Schmidt number $Sc$ and the turbulent Schmidt number $Sc_t$ are set to 1. The emission rate of the passive scalar is used to specify the source term, $S_\phi$.
\section{Model setup}
In this section, the boundary conditions, especially for the inlet and ground surface are discussed in detail. The map for variable aerodynamic roughness length is introduced to include the effects of the surface structures in the complex terrain. Finally, the method to generate high-quality mesh for the complex terrain is described. 

\subsection{Boundary conditions}
\label{sec:BCs}

In buoyantSimpleFoam, boundary conditions and initial values for the following variables are required: $u$, $\theta$, $k$, $\varepsilon$, $\nu _t$, and $p_\textup{rgh}$. For $p$ and $\alpha _t$, their values are directly calculated from $p_\textup{rgh}$ and $\nu _t$. All of the boundary conditions are summarized in Table \ref{tab:2}.

\begin{table}[htb!]
  \centering
  \captionof{table}{Boundary conditions specified in the CFD model.}
  \scalebox{0.9}{
    \begin{tabular}{ccccc}
      \hline
      Variable         & Inlet         & Outlet        & Ground        & Top            \\ \hline
      $u$              & fixed value   & zero gradient & fixed value   & fixed value    \\
      $\theta$         & fixed value   & zero gradient & fixed value   & fixed gradient \\
      $k$              & fixed value   & zero gradient & zero gradient & fixed value    \\
      $\varepsilon$    & fixed value   & zero gradient & wall function & fixed value    \\
      $\nu _t$         & zero gradient & zero gradient & wall function & zero gradient  \\
      $p_\textup{rgh}$ & zero gradient & fixed value   & zero gradient & zero gradient  \\
      $p$              & calculated    & calculated    & calculated    & calculated     \\
      $\alpha _t$      & calculated    & calculated    & calculated    & calculated     \\
      $\phi$           & zero gradient & zero gradient & zero gradient & zero gradient  \\ \hline
    \end{tabular}
  }
  \label{tab:2}
\end{table}
\subsubsection{Inlet boundary conditions}
As the model inputs, the vertical distributions of $u$, $\theta$, $k$, $\varepsilon$ at the domain inlet need to be specified to constrain the model. Based on the Monin–Obukhov similarity theory, velocity $u$ as a function of height $z$ and the Monin-Obukhov length $L$ is calculated as the general form \citep{panofsky1984,Stull1988,Wallace2006}:

\begin{equation}
  u(z)=\frac{u_*}{\kappa}\left[ \ln \left( \frac{z}{z_0}\right)+a \frac{z}{L}\right]
\end{equation}
\label{eqn:inlet_u}
Similarly, based on the ground potential temperature $\theta_0$, the potential temperature profile is calculated as:

\begin{equation}
  \theta (z)=\frac{T_*}{\kappa}\left[ \ln \left( \frac{z}{z_0}\right)+b \frac{z}{L} \right]+\theta_0
\end{equation}
\label{eqn:inlet_theta}
The friction velocity $u_*$ is defined as:

\begin{equation}
  u_* \equiv \sqrt{\frac{\tau _w}{\rho}}=\frac{\nu _t}{u_*} \frac{\partial u }{\partial z}
\end{equation}
where the wall shear stress $\tau _w = \rho \nu_t \frac{\partial u}{\partial z}$. Based on the similarity theory, $T_*$ is given by:

\begin{equation}
  T_*=\frac{\alpha _t}{u_*} \frac{\partial \theta }{\partial z}
\end{equation}
The 1968 experiments on a Kansas plain suggested the following relationships \citep{Alinot2005}:

\begin{equation}
  a=b=5
\end{equation}

It should be noted that the coefficients $a$ and $b$ are derived for the flat terrain with uniform and small aerodynamic roughness length, and different experiments yield different values. For example, \cite{Wallace2006} suggested 8.1. In this study, we fit the measurements with the wind speed and potential temperature equations from the Monin–Obukhov similarity. The detailed steps will be described in Section \ref{section:fit}.

For $k$ and $\varepsilon$, they are calculated as:

\begin{equation}
  k = \frac{u_*^2}{\sqrt{C_{\mu} }} 
\end{equation}
\begin{equation}
  \varepsilon = \frac{u_*^3}{\kappa z}
\end{equation}

Boundary conditions for $\nu_t$ and $p_\textup{rgh}$ at the inlet are set as zero gradient, which is a common practice. Finally, since $p$ and $\alpha_t$ are directly calculated from their corresponding variables $p_\textup{rgh}$ and $\nu _t$, their boundary conditions on all boundaries are set as ``calculated'' in the OpenFoam code. For example, $p$ is calculated using $p=p_\textup{rgh}+\rho g z$.

\subsubsection{Outlet boundary conditions}
The zero gradient boundary condition is specified for all variables but $p_\textup{rgh}$, where a fixed value boundary condition is used. Under neutral conditions, when the potential temperature is uniform in the vertical direction, $p_\textup{rgh}$ is also uniform. As a common practice, the pressure is set to be a constant value at the outlet \citep{Pontiggia2009,Parente2011_2,Balogh2012}. Under stable and unstable conditions, the vertical profile of $p_\textup{rgh}$ is unknown without any measurements, but it would still be a good estimate to set $p_\textup{rgh}$ the same as in the neutral condition assuming weak compressibility far away from the high-temperature stacks of the coke plant.

\subsubsection{Surface boundary conditions}
The ground surface can be treated as a rough wall when simulating the flow in the ABL. For velocity, a no-slip condition is commonly used in the modeling of viscous flows for all stability classes.

In this study, uniform potential temperature is specified at the bottom surface for different stability classes, which is also used by \cite{Pontiggia2009}. Under neutral condition, the potential temperature within the whole 3D domain should be uniform. For stable conditions, there should be heat fluxes out of the domain at the ground. Due to the lack of measurement data, it is not possible to specify variable heat fluxes on the ground, so uniform potential temperatures are also used.

For $k$, $\varepsilon$, and $\nu _t$, which are coupled together by the $k-\varepsilon$ turbulence model, wall functions are applied to bridge the inner region between the ground and the fully-developed turbulent region. Traditionally, the standard rough wall function based on the sand-grain roughness has been widely used in modeling ABL flows. However, it restricts the height of the first cell. Based on the aerodynamic roughness length while preserving the universal law of the wall, a new aerodynamic-based rough wall model can be derived similarly to that is proposed by \cite{Parente2011_2}. Near the wall, the velocity profiles from Equation \ref{eqn:inlet_u} with or without the $\Psi_m$ term will be almost identical, so the dimensionless velocity $u^+$ for all stability classes is given by:

\begin{equation}
  u^+\equiv\frac{u_c}{u^*}=\frac{1}{\kappa} \ln \left( \frac{z_c}{z_0}\right)=\frac{1}{\kappa} \ln \left( E' z^+\right)
\end{equation}
where $E'= \nu /(z_0 u_*)$ and $z^+= z_c u_*/\nu$. The subscript $c$ means the value on the first cell above the wall. The wall shear stress is:

\begin{equation}
  \tau_w = \rho u^{*2} = \rho u^* \frac{u_c}{u^+} = \frac{\rho u^* u_c}{\frac{1}{\kappa}\ln(E'z^+)}
\end{equation}
and it can also be defined as:
\begin{equation}
  \tau_w = \rho \nu_\textup{eff} \frac{u_c}{z_c}
\end{equation}
The effective kinematic viscosity near the wall is:

\begin{equation}
  \nu_\textup{eff}=\nu + \nu_{tc}=\frac{ u^* z_c}{\frac{1}{\kappa}\ln(E'z^+)} = \frac{ z^+ \nu}{\frac{1}{\kappa}\ln(E'z^+)}
\end{equation}
So the new turbulent kinematic viscosity on the wall is updated as:

\begin{equation}
  \nu _{tc} = \nu \left[ \frac{ z^+}{\frac{1}{\kappa} \ln \left( E'z^+\right)}-1\right]
\end{equation}
The turbulent kinetic energy production term $G_k$ in Equation \ref{eqn:keqn} needs to be balanced by the dissipation term $\rho \varepsilon$. By equating them:

\begin{equation}
  G_k = \tau_{w} \frac{\partial u}{\partial z}=\rho \frac{u_*^3}{\kappa z_c} = \rho \varepsilon_c
\end{equation}
$\varepsilon _c$ is obtained and its value is updated on the ground by the following equation:
\begin{equation}
  \varepsilon_c = \frac{u_*^3}{\kappa z_c} = \frac{C_\mu^{3/4}k_c^{3/2}}{\kappa z_c}
\end{equation}

\begin{figure}[hbt!]
  \centering%%trim=left bottom right top  
  \includegraphics[width=0.8\linewidth]{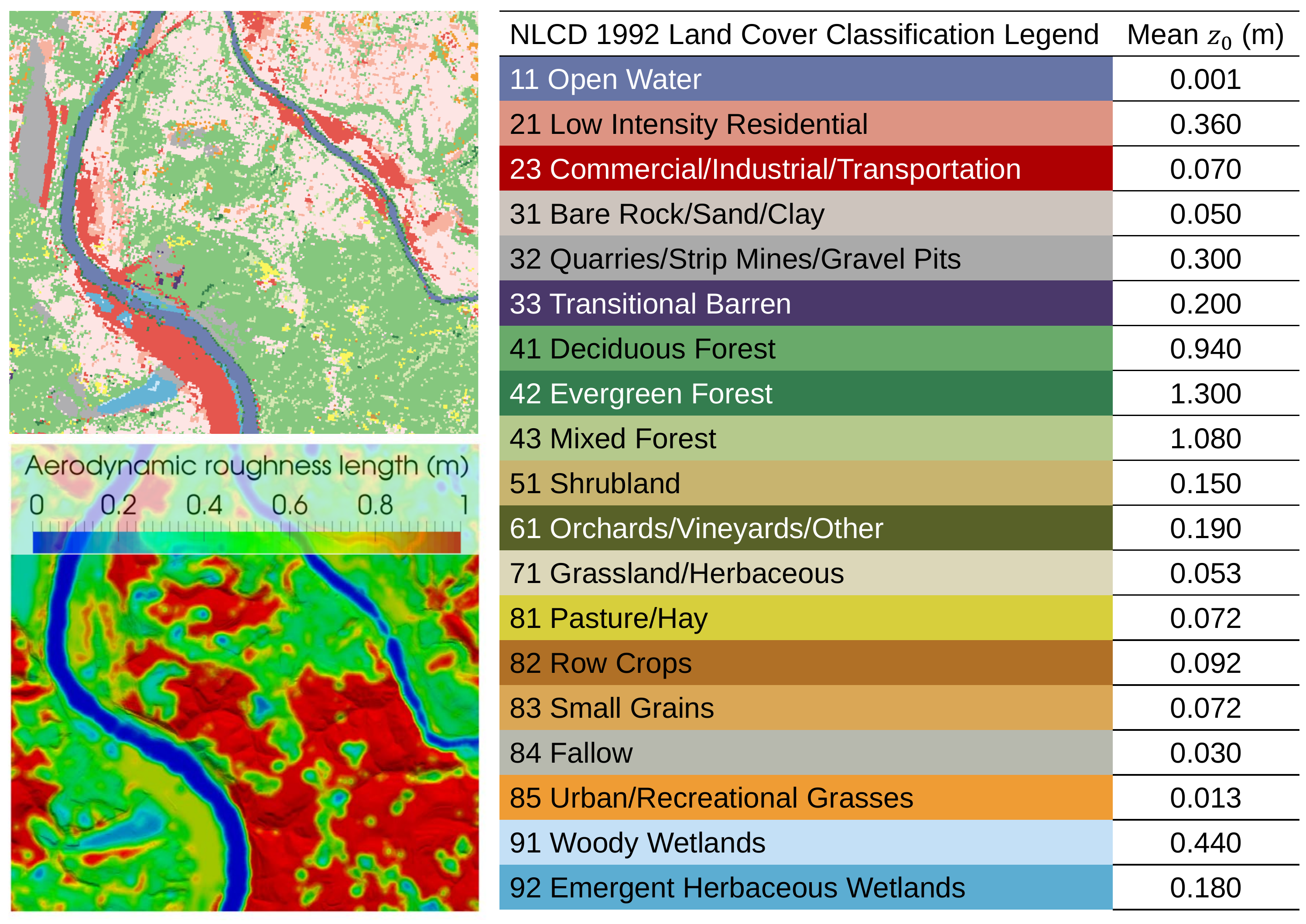}
  \caption{The 1992 land cover data from NLCD (on the upper left) and the table for mean $z_0$ of each land cover class (on the right) as input to AERSURFACE to obtain the roughness map (on the lower left).}
  \label{fig:nlcd1992}
\end{figure}

To use the new rough wall model, the aerodynamic roughness length of the ground surface should be provided. National Land Cover Database (NLCD) is considered when developing the variable aerodynamic roughness length map over the complex terrain that is need for the wall model. The 1992 database with a grid resolution of 30 $m$ is used. The land cover map, which is slightly larger than the study domain is obtained from the Multi-Resolution Land Characteristics (MRLC) Consortium website \citep{MRLC} as shown in Figure \ref{fig:nlcd1992}. Using the AERSURFACE algorithm \citep{Cimorelli2005}, the land cover class can be converted into the aerodynamic roughness length based on the table shown in Figure \ref{fig:nlcd1992}. The average aerodynamic roughness length is based on the inverse-distance weighted geometric mean within a given radius. AERSURFACE recommends a default value of 1 $km$, but the land cover of the study domain may vary a lot within such a radius. The 100 $m$ radius, which is the minimum value AERSURFACE accepts, is used to calculate the average aerodynamic roughness length.

Finally, as for $k_p$ and $p_\textup{rgh}$, the zero gradient boundary condition is commonly used on the ground surface.

\subsubsection{Top boundary conditions}
Based on the vertical profiles of $u$, $k$, and $\varepsilon$, their values on the top boundary are set as fixed using the Dirichlet boundary condition. A uniform fixed gradient of potential temperature is used as the Neumann boundary condition. As for $\nu_t$ and $p_\textup{rgh}$, zero gradient is applied on the top boundary.

\subsubsection{Stack exit boundary conditions}
There are a total of ten stacks on the Coke Plant that are responsible for the majority of the \ch{SO2} emissions. Since the stacks have very small dimensions compared to the overall computational domain, their influence on the flow field was found to be negligible. Therefore, only the stack exits are modeled.  The height of the stacks ranges from 69 to 98 $m$ AGL. The average stack exit temperature is about 500 $K$ and the stack exit vertical velocity is about 5.5 $m/s$. The vertical velocity at each stack exit is set according to measurements while the horizontal velocity is set to zero gradient. In addition to the emissions from the stacks, there is a small amount of \ch{SO2} that is emitted from other locations on the Coke Plant, the exact locations of which are not available. These emissions are classified as fugitive emissions. In the CFD model, the fugitive emissions are specified over the entire plant region and emitted at the ground level.

\subsection{Computational mesh}

\begin{figure}[htb!]
  \centering
  \begin{subfigure}[b]{.4\linewidth}
    \centering
    %%trim=left bottom right top  
    \includegraphics[trim=000 10 10 0,clip,width=0.93 \linewidth]{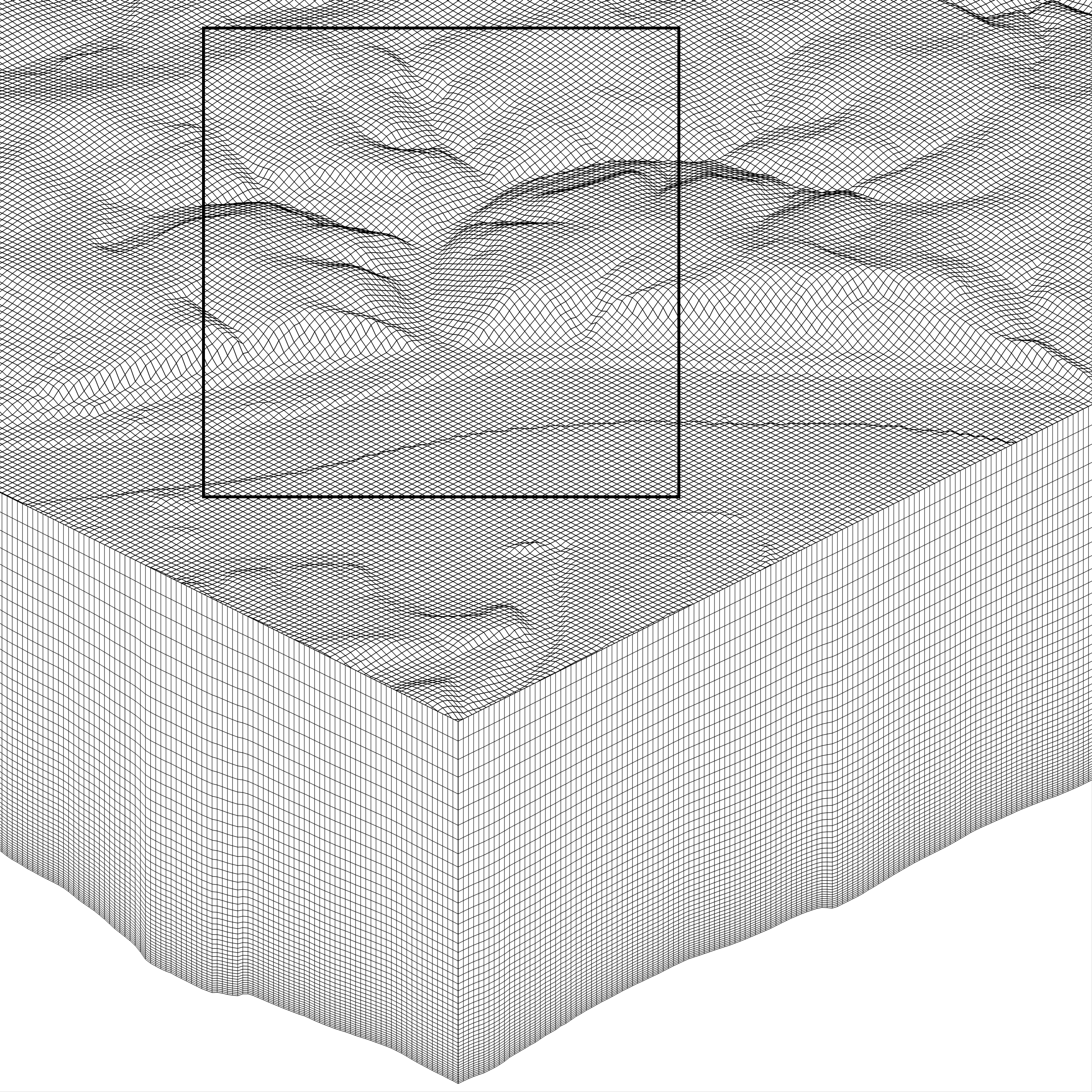}
    \caption{}
    \label{fig:mesh_sw}
  \end{subfigure}%
  \begin{subfigure}[b]{.4\linewidth}
    \centering
    \includegraphics[width=0.93\linewidth]{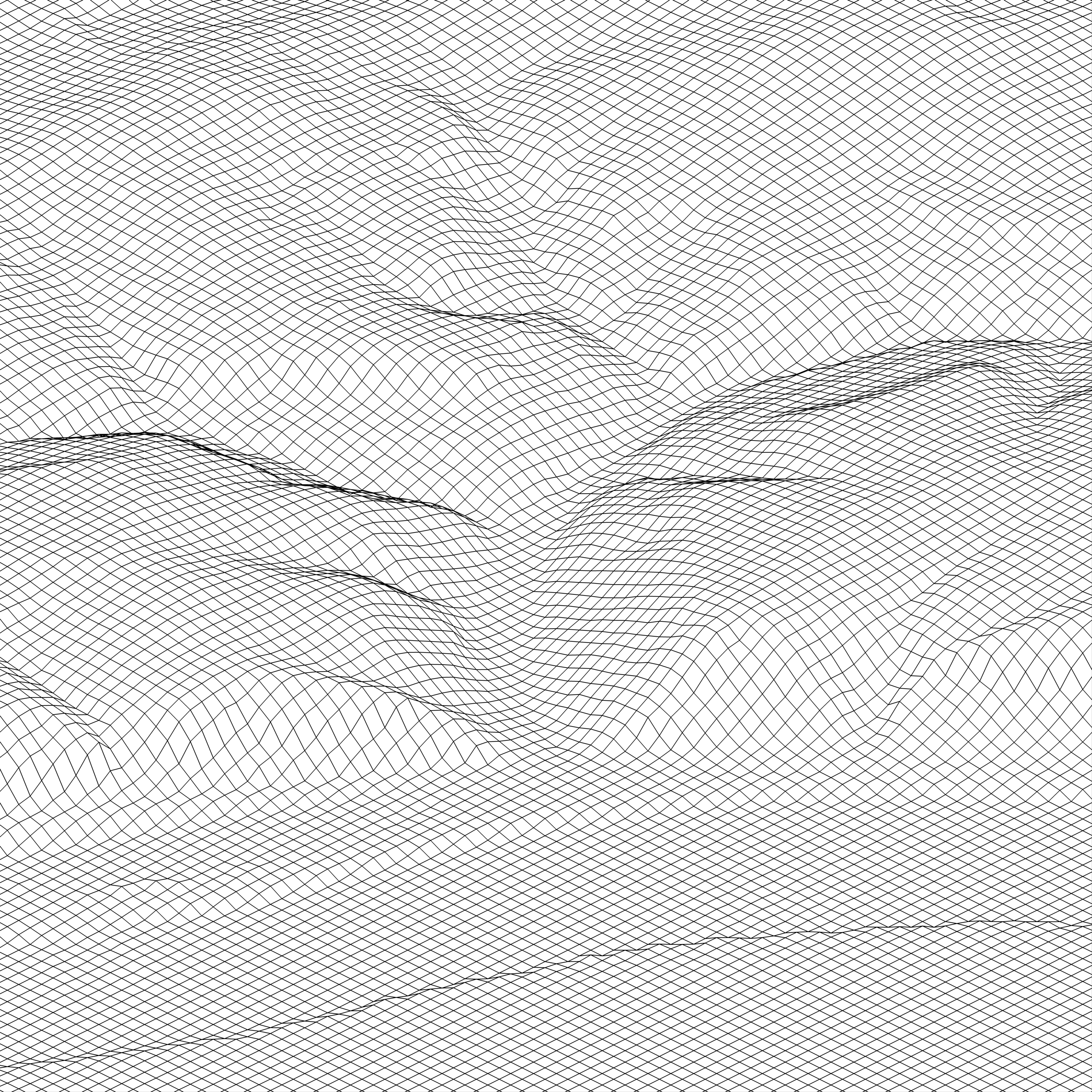}
    \caption{}
    \label{fig:mesh_coke_plant}
  \end{subfigure}%
  \begin{subfigure}[b]{.2\linewidth}
    \centering
    %%trim=left bottom right top  
    \includegraphics[trim=300 0 72 0,clip,width=0.98 \linewidth]{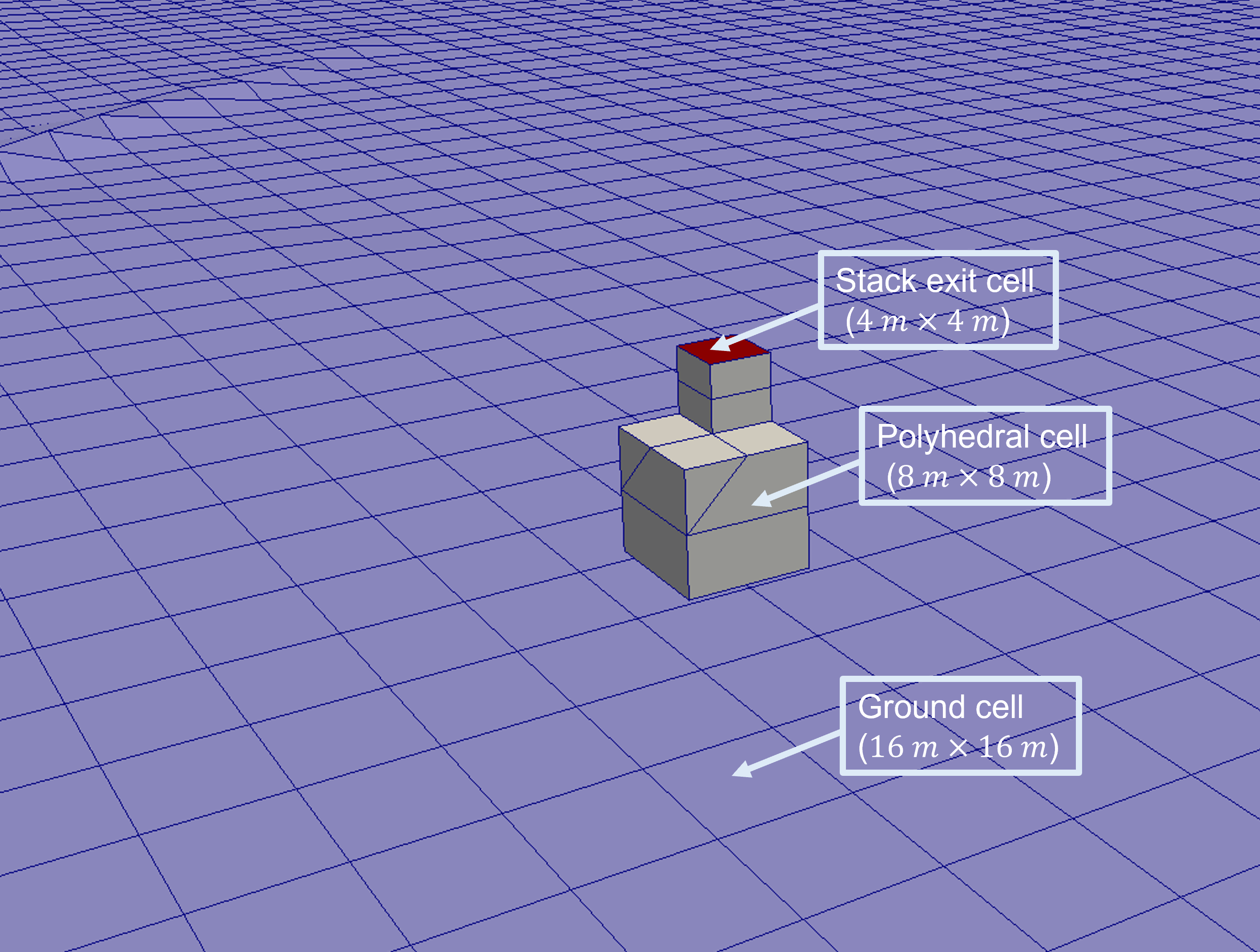}
    \caption{}
    \label{fig:stack_exit_cell}
  \end{subfigure}
  \caption{Views of the computational mesh at (a) southwest corner, (b) hollow near the river, and (c) one stack exit}
  \label{fig:mesh_3}
\end{figure}

The geometry of the terrain surface is obtained from Allegheny County 2006 Contours available on Pennsylvania Spatial Data Access \citep{PASDA2006}. The domain height is set to 1000 $m$. Figure \ref{fig:mesh_3} shows the fully hexahedral mesh created for the study domain. Since the four sides of the domain on the ground are complex curves rather than straight lines, extra steps are needed to generate a good-quality mesh. First, the geometry and mesh generation software, ICEM CFD \citep{ansys}, is used to smoothly extend the terrain surface in four directions to a flat surface. Then, a single block is used to create a fully hexahedral mesh. After that, the snappyHexMesh tool is used to cut off the extensions to restore the original 6.4 $km$ by 6.4 $km$ terrain. Finally, the cells near the stack exits are refined and  the cells just above the exit faces are remove. The final mesh around the stack exits is shown in Figure \ref{fig:stack_exit_cell}. Ideally, only the cells that have the stack exit faces will be removed. However, the local refinement near the stack exits will snap the original hexahedral cells into smaller polyhedral cells, which are grouped together. After removing the stack exit cell, the attached polyhedral cells are also removed. The exposed faces surrounding a stack exit are treated as slip wall.

The total number of hexahedral cells is about 8 million. There are 409 cells in the x-direction (west to east), 409 cells in the y-direction (south to north), and 48 cells in the z-direction (ground surface to top). Each cell has a length and width of about 16 $m$. The height of the first cell above ground is 4 $m$. An expansion ratio of 1.057 is used to increase cell height in the vertical direction. The inlet profiles are set based on the equations discussed in Section \ref{sec:BCs}. However, since the terrain is complex, there are different elevation levels of the ground surface at the inlet as shown in Figure \ref{fig:mesh_sw}. The height $z$ in the equations for different inlet boundary profiles is set to be the local height above ground level (AGL).

\section{Methodology to construct inlet boundary profiles}

\begin{figure}[htb!]
  \centering
  \includegraphics[width=0.6\linewidth]{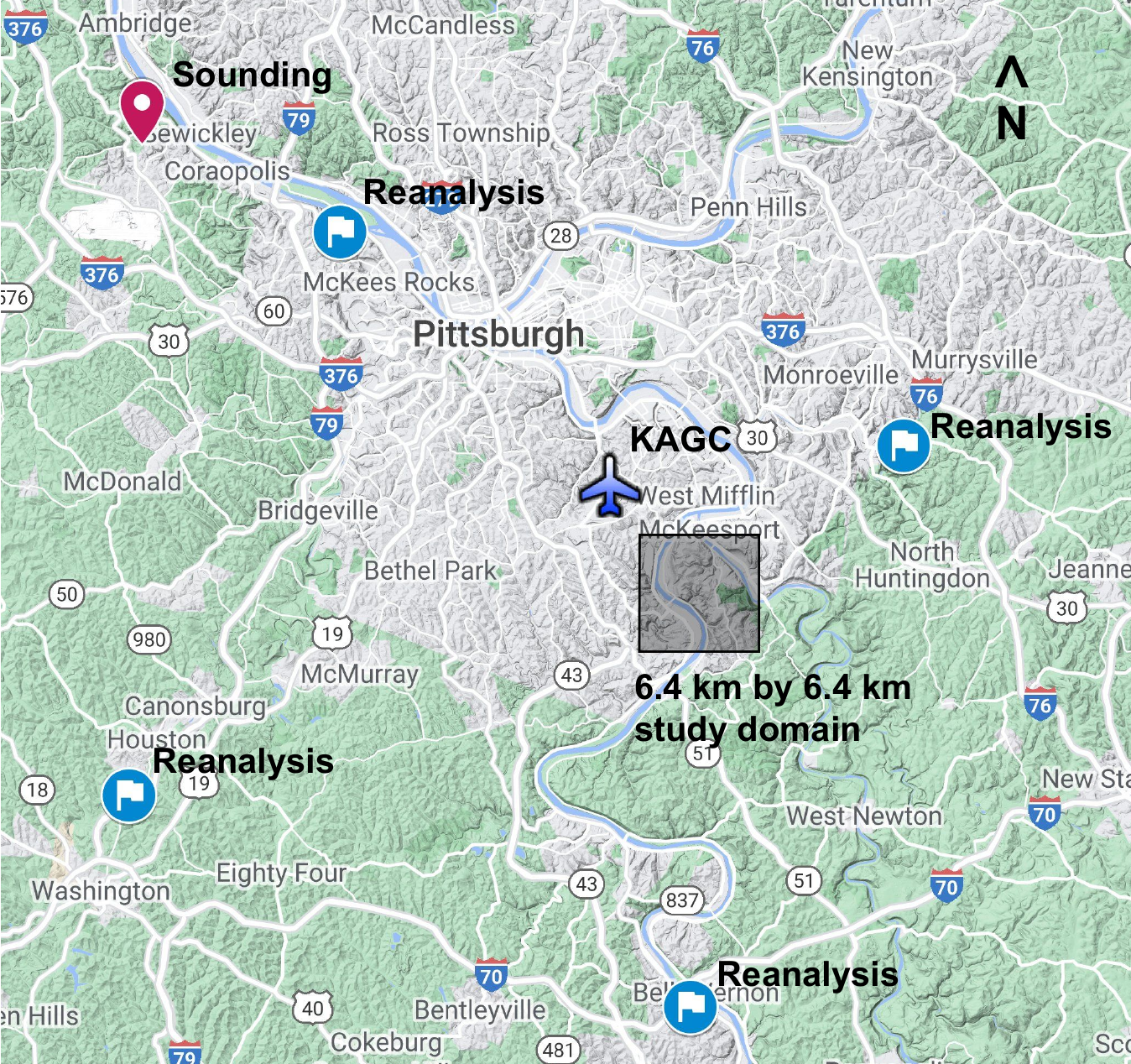}
  \caption{Locations of different meteorological data sources outside the study domain. These sources are considered to be permanent, and they are used to generate vertical boundary profiles to be specified at the inlet of the domain.}
  \label{fig:domain_wind_data}
\end{figure}

The boundary profiles at the inlet of the domain are key to an accurate simulation. This section explains the development of a curve-fitting method to generate boundary profiles at inlet based on MOST with inputs from ground-level and vertically distributed weather data. The locations of different meteorological data sources outside the study domain are shown in Figure \ref{fig:domain_wind_data}.

\subsection{Meteorological data sources}
\subsubsection{Sounding data}
The sounding data are reported twice at 00Z and 12Z in Zulu time, which correspond to 19:00 and 07:00 US EST. The Skew-T plots generated from the sounding data are used to determine the stability class of the atmosphere. The neutral condition is easily determined if the temperature profile is parallel to any of the dry adiabatic lapse rate lines. Strictly speaking, the stable class is identified when the environmental lapse rate is less than the moist adiabatic lapse rate. Since the moisture in the air is not modeled, and the air in the CFD model is dry, the stable class is identified when the environmental lapse rate is less than the dry adiabatic lapse rate.

\subsubsection{Surface weather observations}
\begin{figure}[htb!]
  \centering
  \includegraphics[width=0.5\linewidth]{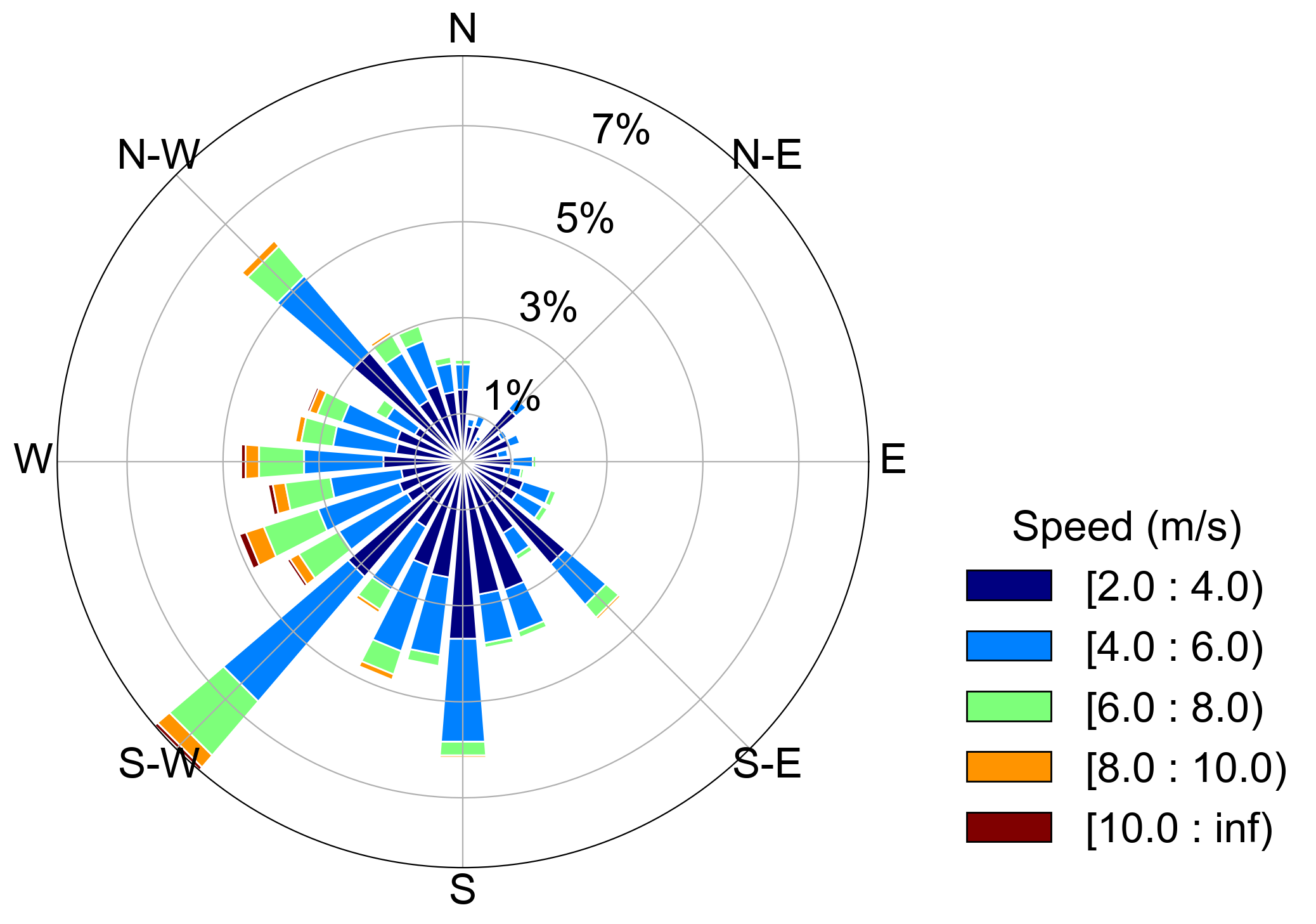}
  \caption{Wind rose at KAGC including all wind data collected from 1945 to 2019. The calm wind speed is below 2 $m/s$.}
  \label{fig:windrose_KAGC}
\end{figure}

In addition to the sounding data, surface weather observations collected by National Weather Service are available. The nearby Allegheny County Airport (KAGC) reports wind speed and direction that are measured at 10 $m$ AGL. A wind rose is generated using data from KAGC from the year 1945 to 2019. As shown in Figure \ref{fig:windrose_KAGC}, the dominant wind direction is from the southwest, and the most common wind speed is from 4 to 6 $m/s$. Therefore, conditions with southwest wind are considered in the present work.

\subsubsection{Reanalysis data}

The National Centers for Environmental Prediction/National Center for Atmospheric Research (NCEP/NCAR) Reanalysis 1 project provides reanalysis data at 17 pressure levels (from 1000 $mb$ to 10 $mb$) for every 3 hours from 00Z to 21Z  \citep{Mesinger2006}. Reanalysis data are generated by utilizing available observations over the period being analyzed. The data provide a dynamically consistent estimate of the atmosphere. As shown in Figure \ref{fig:domain_wind_data}, there are 4 locations surrounding the study domain where the reanalysis data in the vertical direction are available, and their distances to the domain range from 10 $km$ to 30 $km$. The mean vertical profiles from the 4 locations are used in the present work. The mean difference over several months between the reanalysis data and sounding data is about $20\%$ in wind speed and about 1 $K$ in temperature. When sounding data is not available, the reanalysis can be used to establish the vertical profiles of the ABL.

\subsection{Curve-fitting method}
\label{section:fit}
\begin{figure}[htb!]
  \centering
  \includegraphics[width=0.9\linewidth]{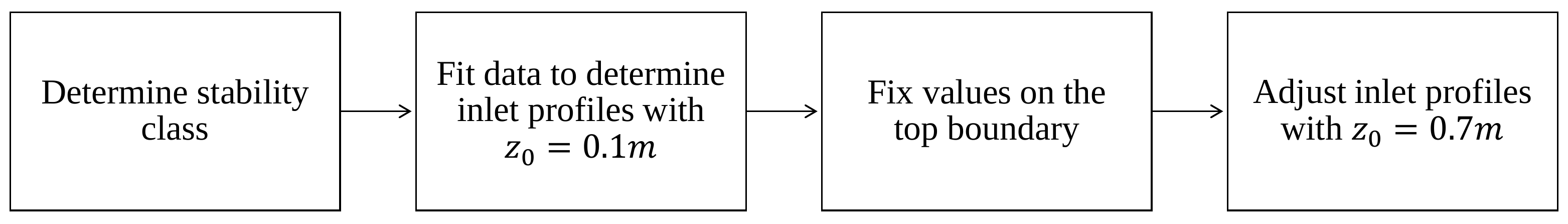}
  \caption{Flowchart of the curve-fitting procedure used in the present work to obtain boundary conditions at the inlet of the study domain.}
  \label{fig:curve_fitting_flowchart}
\end{figure}

In order to provide a consistent estimate of the atmosphere by taking into account data measured from multiple height levels, a curve-fitting method is developed. The curve-fitting method aims to minimize the non-linear least-square errors in the theoretical equations from MOST. To this end, the optimal parameters in the equations are determined. The procedure is summarized in the flowchart shown in Figure \ref{fig:curve_fitting_flowchart}.

At the beginning, the stability class of the ABL is determined from the vertical temperature profile of the sounding if it is available. Otherwise, the temperature profile from reanalysis is used. The sounding data is also preferred over the reanalysis data for the vertical wind speed profile. The ``curve$\_$fit" function from the ``scipy" package \citep{sciPy2020} is employed to fit the theoretical equations. The sigma values for the KAGC data and sounding data are set to 0.1 and 0.5 respectively. The KAGC data are reported by sensors on a fixed height at 10 $m$ AGL, they have less uncertainty. As wind speed increases with height, the associated uncertainty is larger \citep{Langreder2016}. Also, the surface layer where the similarity theory applies is low to the ground. Thus, a smaller sigma value is chosen for the KAGC data. As for the sounding data, they represent the instantaneous condition of the atmosphere, so more uncertainty is set for them. To minimize the local effects from the sounding location, only data points with height above 500 $m$ with respect to the height of the Monongahela River surface inside the study domain are used.

For the neutral condition, the method only needs to fit the velocity equation. Since the KAGC airport and the sounding location have similar aerodynamic roughness lengths estimated as 0.1 $m$, the velocity equation is fitted by setting $z_0$ to 0.1 $m$. In this way, the initial vertical wind speed distribution is obtained. The velocity at the top of the boundary layer (1000 $m$) is determined by plugging the fitted parameters into Equation \ref{eqn:inlet_u}. The aerodynamic roughness length is then changed to 0.7 $m$, which is the average value at the inlet of the study domain. With the new aerodynamic roughness length at the inlet and the fixed top velocity, a new set of parameters in the equations, such as $u_{*}$ is obtained. The reason for changing the roughness length is to adjust the wind speeds near the ground at the domain inlet. For the stable condition, the method simultaneously fits data into the inlet velocity equation and potential temperature equation to find the optimal values of all parameters, such as $T_*$, the coefficients $a$, $b$.

\begin{figure}[htb!]
  \centering
  \begin{subfigure}{0.5\linewidth}
    \centering
    %%trim=left bottom right top  
    \includegraphics[ width=0.9\linewidth]{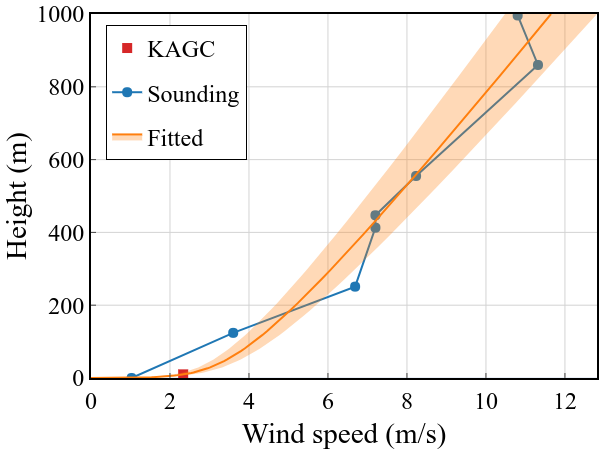}
    \caption{}
    \label{fig:stable1_Uprofiles_new}
  \end{subfigure}%
  \begin{subfigure}{0.5\linewidth}
    \centering
    \includegraphics[ width=0.9\linewidth]{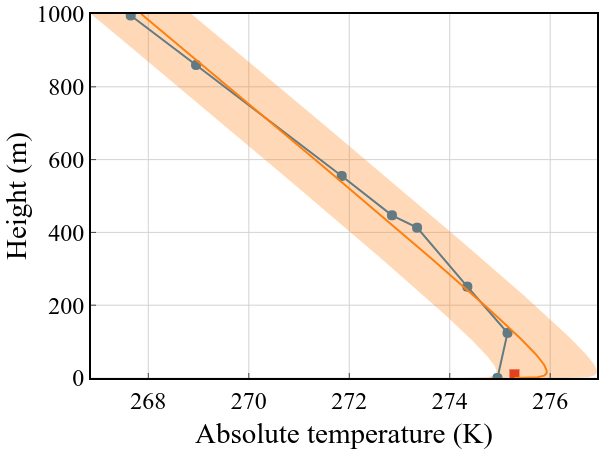}
    \caption{}
    \label{fig:stable1_Tprofiles_new_nolegend}
  \end{subfigure}
  \caption{An example of curve-fitting results obtained for (a) wind speed and (b) absolute temperature for one of the stable cases referred to as stable case 1.}
  \label{fig:stable1}
\end{figure}

\begin{figure}[htb!]
  \centering
  \begin{subfigure}{0.5\linewidth}
    \centering
    %%trim=left bottom right top  
    \includegraphics[width=0.9\linewidth]{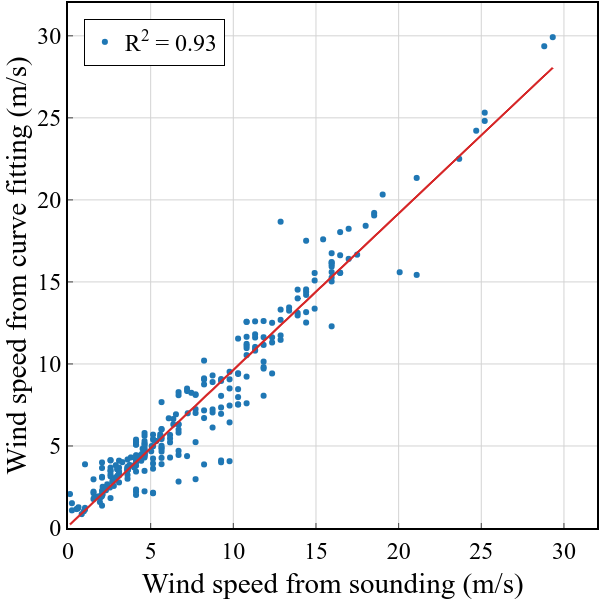}
    \caption{}
    \label{fig:allSpeedR2_Stable_2019_03_new}
  \end{subfigure}%
  \begin{subfigure}{0.5\linewidth}
    \centering
    \includegraphics[width=0.9\linewidth]{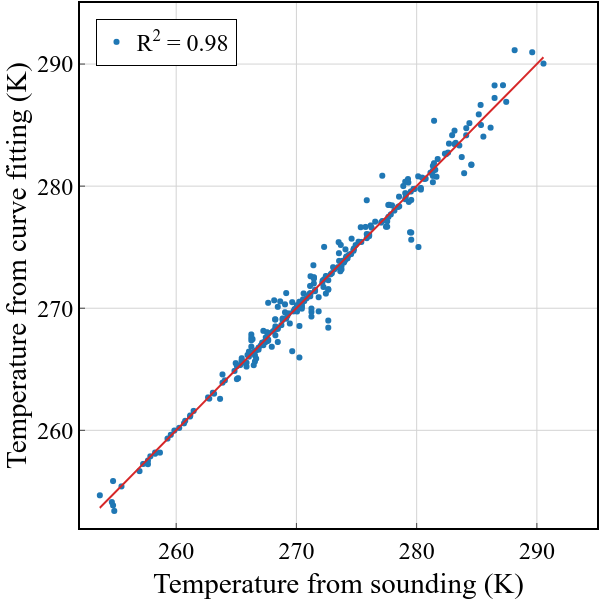}
    \caption{}
    \label{fig:allTempR2_Stable_2019_03}
  \end{subfigure}
  \caption{Evaluation of fitted profiles versus measurements from atmospheric soundings for (a) wind speed and (b) air temperature over a month.}
  \label{fig:R2_Stable_2019_03}
\end{figure}
Figure \ref{fig:stable1} shows an example of the result from the curve-fitting method. The “Fitted” curve for velocity is shown with $\pm 10\%$ range of the mean value. The “Fitted” curve for temperature is shown with $\pm$ 1 $K$ range of the mean value. The temperature profiles are shown in terms of the absolute air temperature, not the potential temperature.  To test the performance of the curve-fitting method, the wind speed data within a month are fitted into a theoretical wind profile under neutral conditions. The $R^2$ value is about 0.7 when compared to the sounding data. As shown in Figure \ref{fig:R2_Stable_2019_03}, the curve-fitting method yields a much better $R^2$ value of 0.93 for wind speed in case of stable conditions. As for temperature, the $R^2$ value is 0.98. Since the instantaneous wind profiles from sounding do not always have a logarithmic shape, if these non-logarithmic groups are ignored, the $R^2$ value will be higher, especially when fitting data under the neutral condition.

\section{Results and discussion}

The CFD model uses the RANS-based turbulence approximations, which are suitable for simulating the mean flow behavior. Using inlet boundary conditions calculated from different meteorological conditions, we have used the model to predict wind development over the study domain under quasi steady-state conditions. Quasi steady-state here means that the change in wind direction and speed are relatively small over a short period that they can be assumed constant during that period. In the present work, this period is defined to be three hours. If the weather conditions change dramatically during this period, the measured data will show significant variation in wind speed, wind direction, and possibly air temperature. The steady-state model will not be able to predict these changes, and a transient model will be necessary. A steady-state simulation requires that the boundary conditions are kept unchanged. Therefore, for model validation, three criteria are established to select cases for simulations. (1) The vertical temperature profile from the Skew-T plot needs to show a clear stability class. (2) The weather condition should remain quasi steady-state for at least three hours. (3) The curve-fitting method should provide satisfactory vertical profiles when compared with available meteorological data sources. 

\subsection{Measurements within the study domain}

\begin{figure}[htb!]
  \centering
  \begin{subfigure}{0.11\linewidth}
    \centering
    \includegraphics[trim=1400 0 0 500,clip,width=\linewidth]{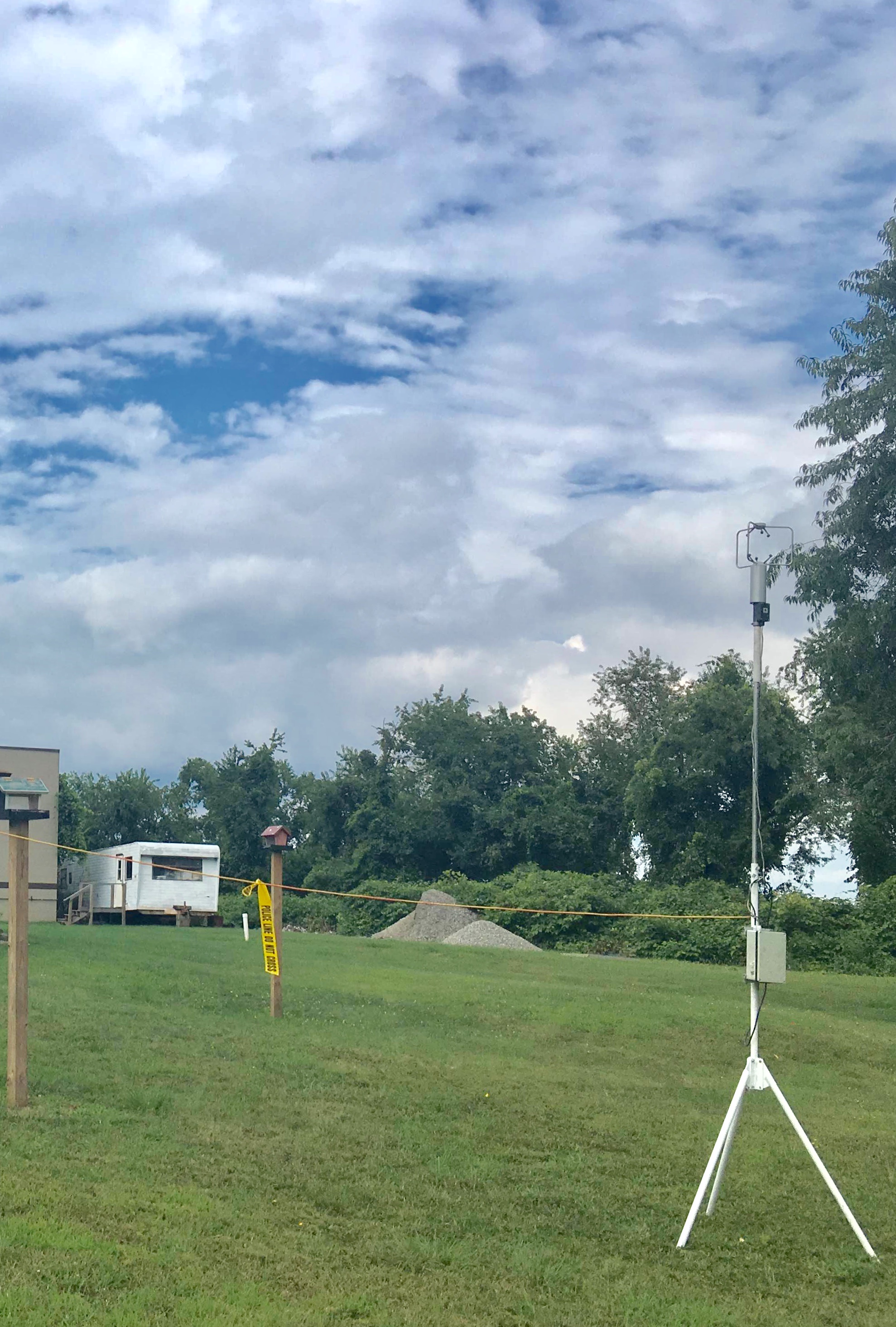}
    \caption{}
    \label{fig:pole}
  \end{subfigure}%
  \begin{subfigure}{0.53\linewidth}
    \centering
    %%trim=left bottom right top  
    \includegraphics[width=0.9\linewidth]{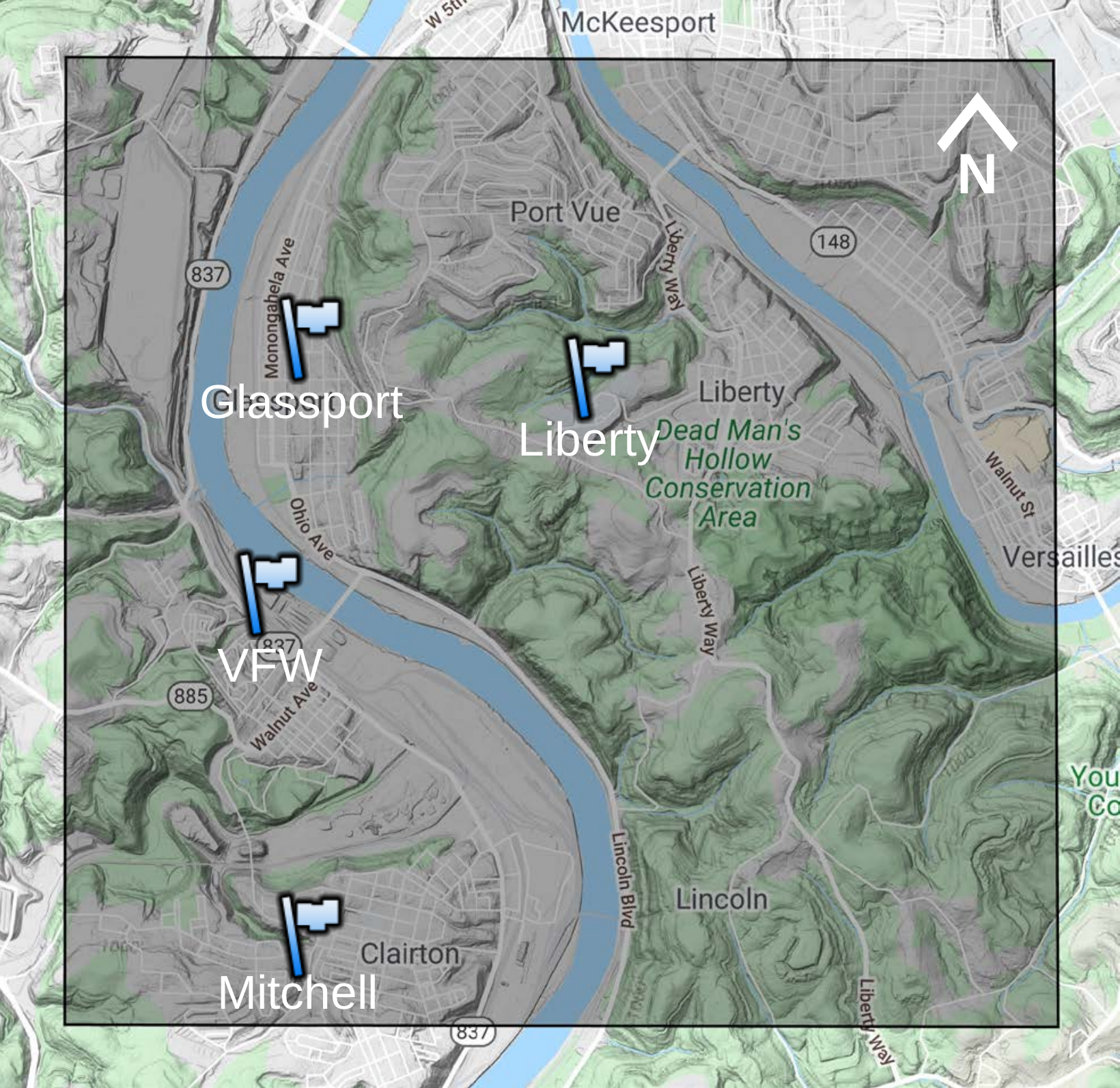}
    \caption{}
    \label{fig:domain_sensor_new}
  \end{subfigure}
  \caption{(a) Photo of the sonic anemometers used in the present work and (b) locations of four anemometers installed in the study domain to evaluate the model.}
  \label{fig:domain_sensor_pole}
\end{figure}

For model validation, measurements inside the study domain are crucial. Four ultrasonic anemometers from RM Young (shown in Figure \ref{fig:pole}) were deployed to cover the variation in elevation and terrain features, so that the sensor network is able to detect interesting flow patterns within the domain. These anemometers can detect a wide range of wind speed ($0-40$ $m/s$) with high accuracy (error: $\pm 3\%$) and quick response time ($<$ 1 $s$). As shown in Figure \ref{fig:domain_sensor_new}, the sensors are labeled as Mitchell, VFW, Glassport, and Liberty based on their locations. At the Liberty site, \ch{SO2} concentrations are recorded continuously. The model of the \ch{SO2} monitor is the Model T100 \ch{SO2} Analyzer from Teledyne API, which uses ultraviolet fluorescence to measure \ch{SO2} in the ranges of $0-50$ $ppb$ and $0-20000$ $ppb$ with a detection limit of 0.4 $ppb$. The precision of the monitors is 0.5$\%$ of reading above 50 $ppb$.

Table \ref{tab:3} provides a description of each sensor location. All the sensors are located in the 17T UTM Zone. Note that the elevation shown in Table \ref{tab:3} is the height of the ground at the location with respect to the lowest point in the domain. The height AGL is the height of the sensor with respect to the local ground. Sensors are mounted at AGL height varying from 2 $m$ at the Mitchell location to 16 $m$ at the Liberty location. Aerodynamic roughness length at each sensor location is also provided in the Table \ref{tab:3}.

\begin{table}[htb!]
  \centering
  \captionof{table}{Descriptions of wind measurement sites within the study domain that are used for model validation.}
  \scalebox{0.9}{
    \begin{tabular}{cccccc}
      \hline
      Location       & KAGC    & Liberty & Mitchell & Glassport & VFW     \\ \hline
      Easting (m)    & 591115  & 596165  & 594482   & 594056    & 593937  \\
      Northing (m)   & 4467441 & 4464307 & 4460663  & 4464732   & 4462931 \\
      Elevation (m)  & 160     & 121     & 92       & 18        & 18      \\
      Height AGL (m) & 10      & 16      & 2        & 8         & 4       \\
      $z_0$ (m)      & 0.10    & 0.52    & 0.59     & 0.68      & 0.44    \\ \hline
    \end{tabular}
  }
  \label{tab:3}
\end{table}

\subsection{Model predictions of wind speed}
The predictions of the CFD model are compared with wind speeds reported at four locations inside the study domain. A total of five simulations under different meteorological conditions are performed. While specifying the boundary conditions at the inlet, it is assumed that the wind direction does not change with height. Wind direction reported by the Liberty monitor is used at the inlet. For the CFD predictions, the results are extracted at the exact location corresponding to the measurement location of the four wind sensors within the study domain. All simulations cases have southwest wind directions but cover different wind speeds in different seasons. The steady-state simulation results are compared with measurements from different locations. 
\subsubsection{Neutral thermal conditions}

\begin{figure}[htb!]
  \centering
  \begin{subfigure}{.5\linewidth}
    \centering
    \includegraphics[width=0.9\linewidth]{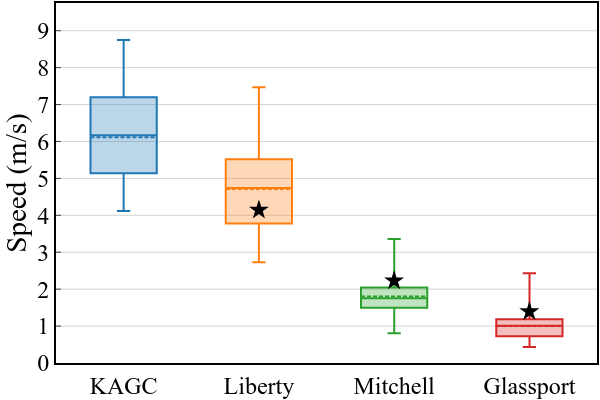}
    \caption{}
    \label{fig:neutral1_box}
  \end{subfigure}%
  \begin{subfigure}{.5\linewidth}
    \centering
    \includegraphics[width=0.9\linewidth]{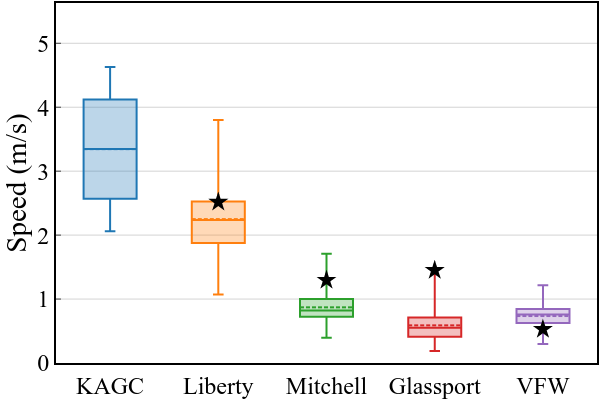}
    \caption{}
    \label{fig:neutral2_box}
  \end{subfigure}
  \begin{subfigure}{.5\linewidth}
    \centering
    \includegraphics[width=0.9\linewidth]{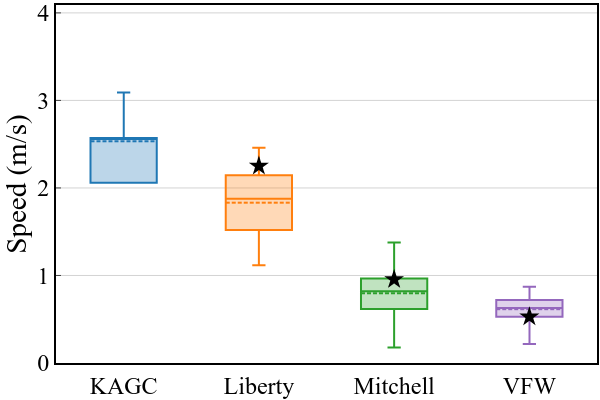}
    \caption{}
    \label{fig:neutral3_box}
  \end{subfigure}

  \caption{Comparison between CFD predictions and measurements under neutral ABL. $\bigstar$: CFD prediction at the monitor. Each box contains measurements during the simulated 3-hour period. The dash line in each box indicates the mean value. (a) Neutral case 1. (b) Neutral case 2. (c) Neutral case 3.}
  \label{fig:neutral123result}
\end{figure}

Under the neutral conditions, three cases are selected corresponding to three different wind speeds shown in Figure \ref{fig:neutral123result}.  The measured data within the $\pm 1$ hour range of the simulated time period are shown as box plots. Each box contains 3 hours of wind speed data. The maximum value, 75th percentile, median, 25th percentile and the minimum value from measurements are shown with each box. The dash line in each box is the mean value, which is close to the median value. The wind speed data at KAGC are shown as a reference. The KAGC location is outside the study domain. Measurements of wind speed at this location are used to construct boundary condition profiles at the inlet of the domain.

The KAGC site has the highest elevation (160 $m$) and is located at a relatively high level above ground. In addition, $z_0$ at KAGC is only 0.1 $m$, which is the smallest among all sites reported in Table \ref{tab:3}.  As a result, the measured wind speed at the KAGC site is always the highest. The Liberty site has an elevation of 121 $m$ and the measured wind speed slightly lower than that of the KAGC site. The average error is within $10\%$ at the Liberty site where relative fast wind speeds are recorded compared to other sites within the domain. As the height decreases, the measured wind speeds at Mitchell become smaller, and further decreases to the lowest level at VFW. Note that the VFW sensor is mounted 4 $m$ above the ground while the Mitchell sensor is mounted at 2 $m$ above the ground. So one might expect a lower wind speed at the Mitchell sensor because of its proximity to the ground. However, the elevation of the Mitchell sensor is higher than the VFW sensor. Both the measurements and predictions show that the wind speed is higher at the Mitchell site. This is an encouraging trend predicted by the model. The predicted wind speeds at Liberty, Mitchell, and VFW are close to the measurements. The predicted wind speed at Glassport is higher than the measurements, especially when the measured wind speed is very low. The model predicts the correct trend of decreasing wind speed from Liberty to Mitchell, and to VFW. In general, the CFD model makes reasonable predictions under neutral conditions. 

\subsubsection{Stable thermal conditions}

\begin{figure}[htb!]
  \centering
  \begin{subfigure}{0.5\linewidth}
    \centering
    %%trim=left bottom right top  
    \includegraphics[width=0.9\linewidth]{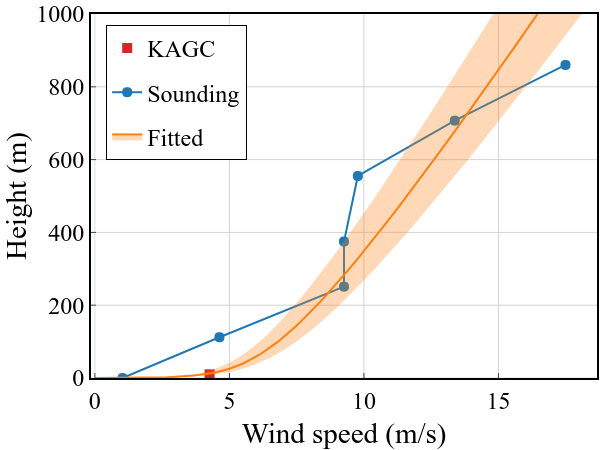}
    \caption{}
    \label{fig:stable2_Uprofiles_new}
  \end{subfigure}%
  \begin{subfigure}{0.5\linewidth}
    \centering
    \includegraphics[width=0.9\linewidth]{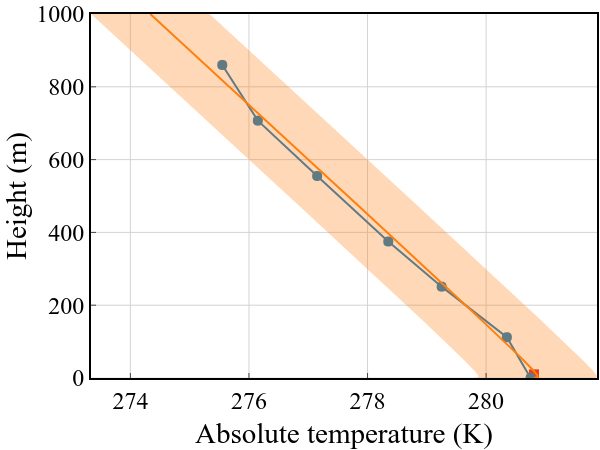}
    \caption{}
    \label{fig:stable2_Tprofiles_new_nolegend}
  \end{subfigure}
  \caption{Curve-fitting results for (a) wind speed and (b) absolute temperature in stable case 2.}
  \label{fig:stable2}
\end{figure}

Two cases are considered for model validation under stable conditions. The results of the curve-fitting method for wind speed and temperature profiles are shown in Figures \ref{fig:stable1} and \ref{fig:stable2} for these two cases. Stable case 1 has moderate inversion with a height of 100 $m$ as shown in Figure \ref{fig:stable1}. Stable case 2 shows a lightly stable condition with relatively higher wind speeds (Figure \ref{fig:stable2}). 

\begin{figure}[htb!]
  \centering
  \begin{subfigure}{0.5\linewidth}
    \centering
    %%trim=left bottom right top  
    \includegraphics[width=0.9\linewidth]{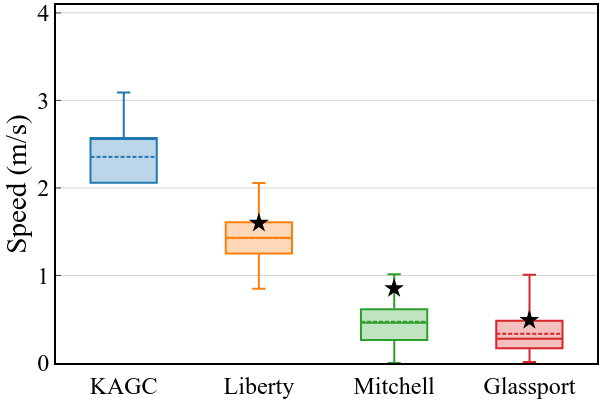}
    \caption{}
  \end{subfigure}%
  \begin{subfigure}{0.5\linewidth}
    \centering
    \includegraphics[width=0.9\linewidth]{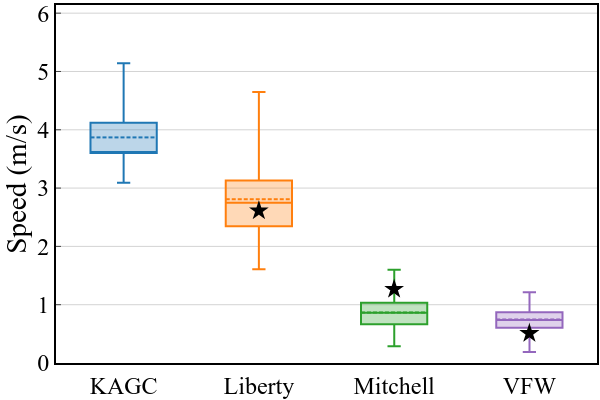}
    \caption{}
  \end{subfigure}
  \caption{Comparisons between CFD predictions and measurements under stable ABL for (a) stable case 1 and (b) stable case 2. $\bigstar$: CFD prediction at the monitor. Each box contains measurements during the simulated 3-hour period. The dash line in each box indicates the mean value.}
  \label{fig:stable_3boxes}
\end{figure}

The comparisons between simulation results and measurements are shown in Figure \ref{fig:stable_3boxes}. Similar to the neutral conditions, the case-to-case trend for the simulations under stable conditions is clear and consistent when compared with the experimental measurements. Therefore, the present model is able to predict wind development in a complex terrain for both neutral and stable classes of the atmosphere. 

\subsubsection{Effects of the complex terrain on flow patterns}
In addition to the comparisons between the CFD predictions and measurements at certain locations for wind speeds, it is also important to understand the effects of the complex terrain on flow patterns. The complex terrain consists of two major parts: irregular topography and variations in land use or surface roughness. The results of different variables based on neutral case 1 are used to qualitatively describe the effects.

\begin{figure}[hbt!]
  \centering
  %%trim=left bottom right top  
  \includegraphics[trim=500 100 200 50,clip,width=0.5\linewidth]{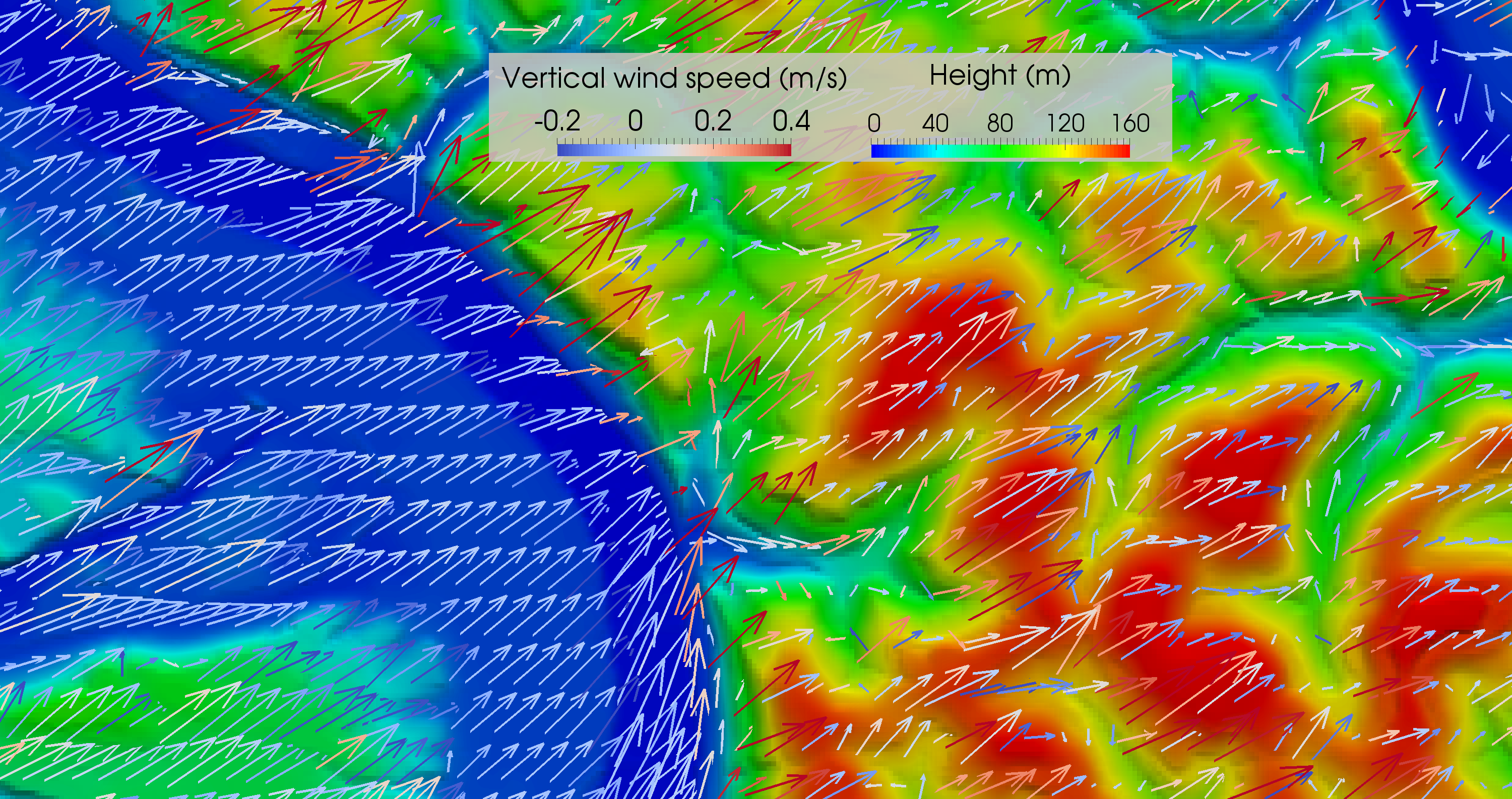}
  \caption{Contours of height above the ground and wind vectors colored by the magnitude of the vertical wind speed.}
  \label{fig:contours_z_vector}
\end{figure}

\begin{figure}[htb!]
  \centering
  \begin{subfigure}{0.24\linewidth}
    \centering
    %%trim=left bottom right top  
    \includegraphics[trim=900 120 890 190,clip,width=1.0\linewidth]{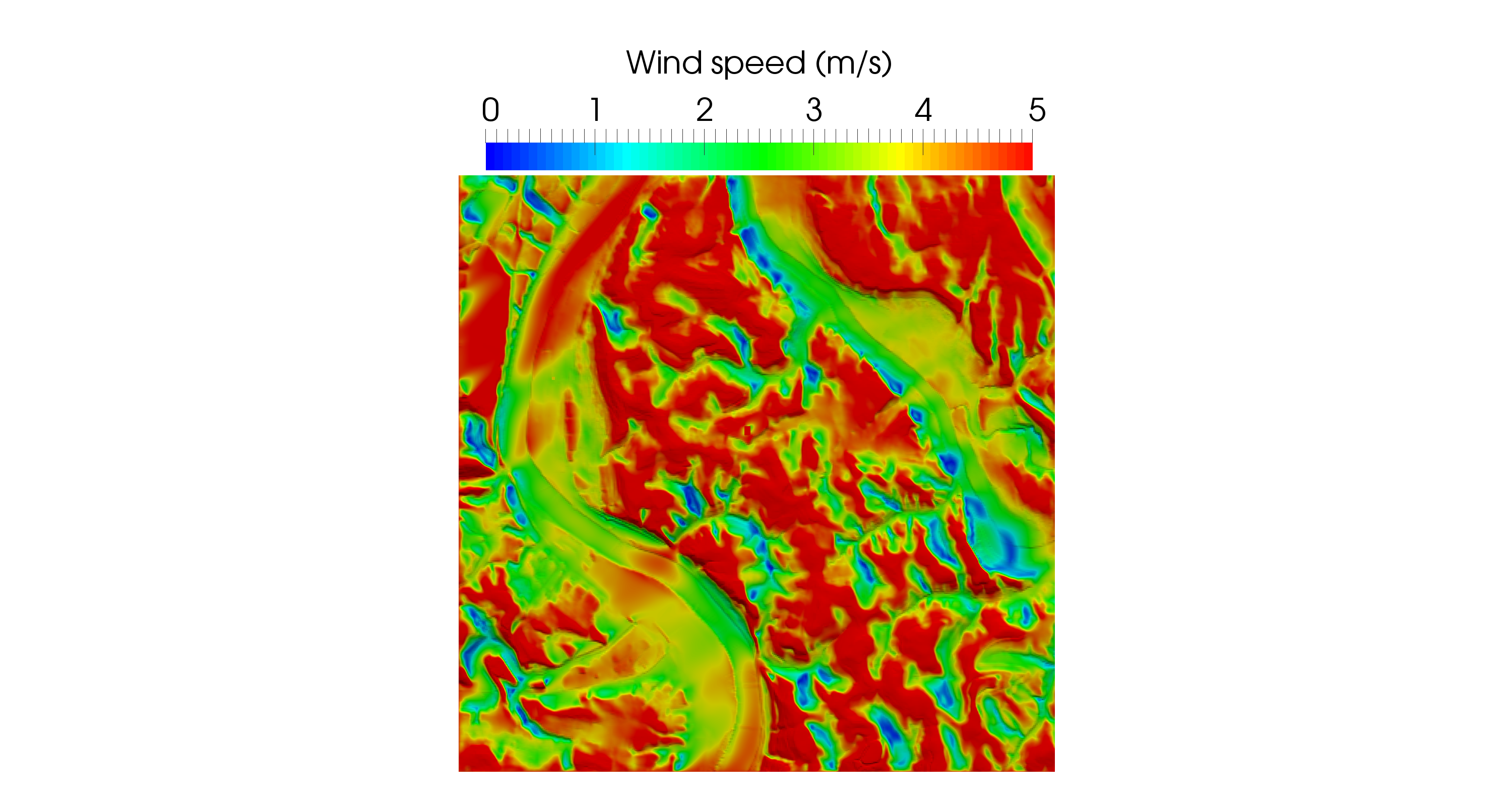}
    \caption{}
    \label{fig:contour_u_10m}
  \end{subfigure}
  \begin{subfigure}{0.24\linewidth}
    \centering
    \includegraphics[trim=900 120 890 190,clip,width=1.0\linewidth]{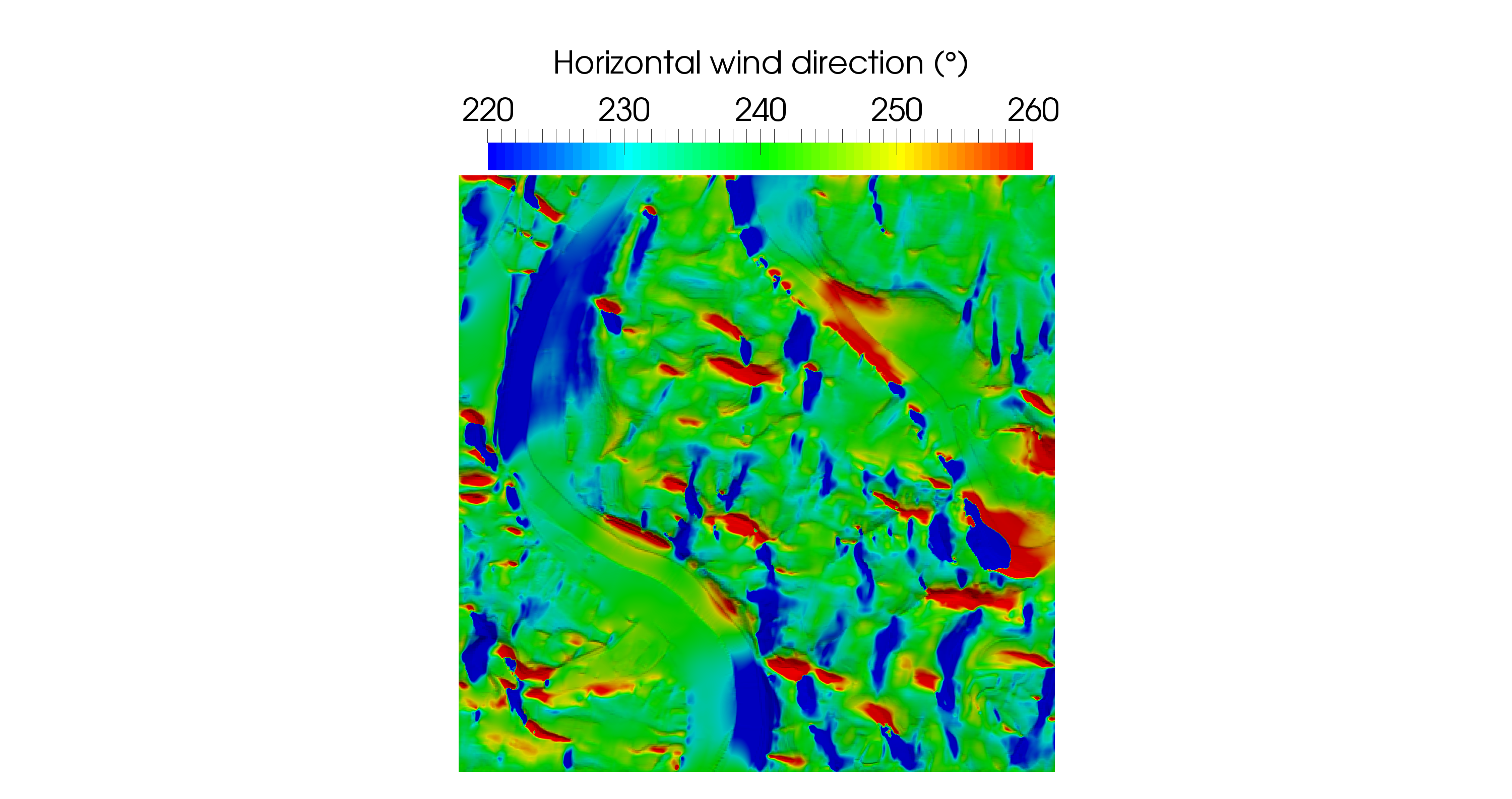}
    \caption{}
    \label{fig:contour_u_dir_10m}
  \end{subfigure}
  \begin{subfigure}{0.24\linewidth}
    \centering
    \includegraphics[trim=900 120 890 190,clip,width=1.0\linewidth]{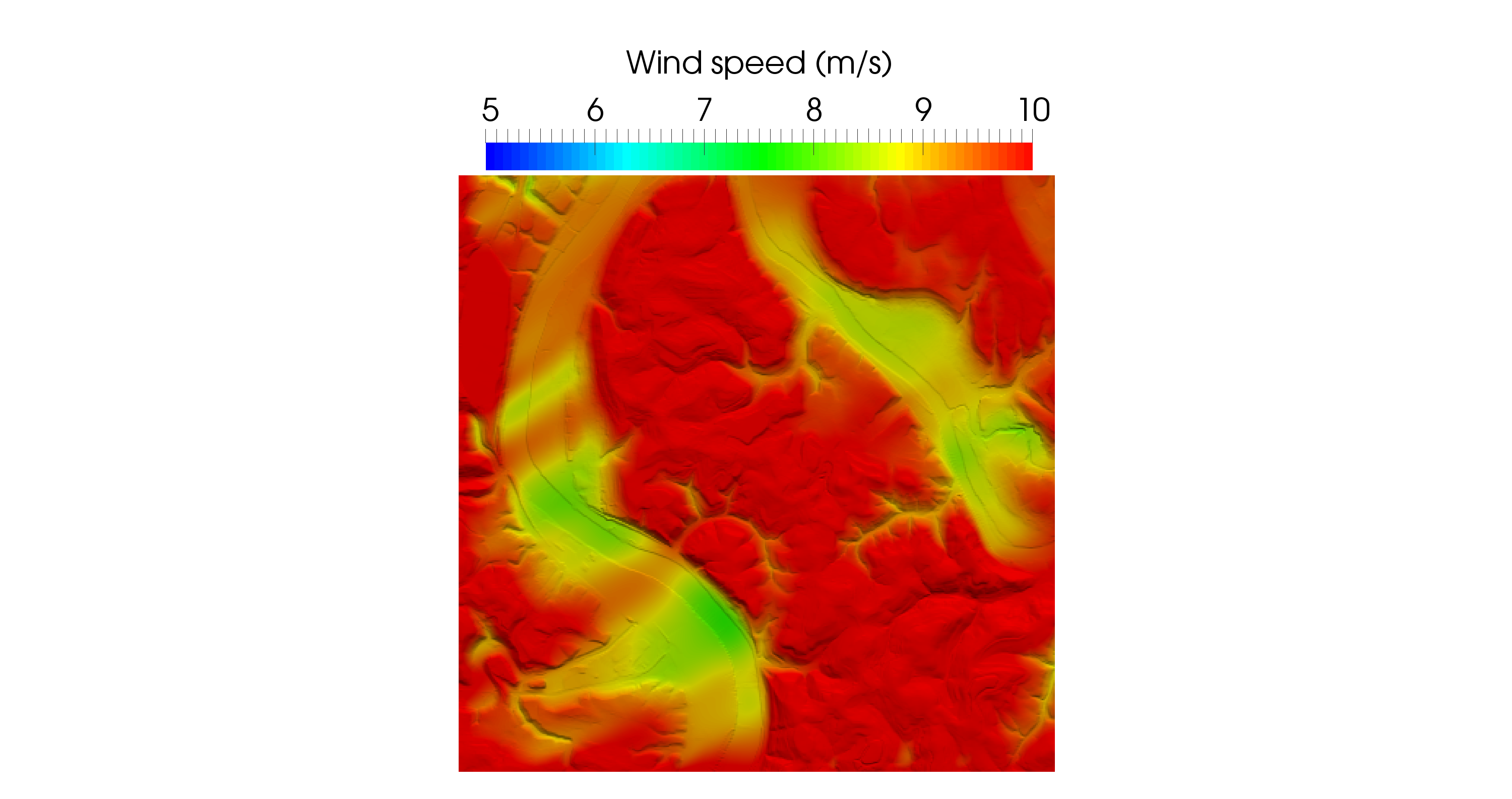}
    \caption{}
    \label{fig:contour_u_200m}
  \end{subfigure}
  \begin{subfigure}{0.24\linewidth}
    \centering
    \includegraphics[trim=900 120 890 190,clip,width=1.0\linewidth]{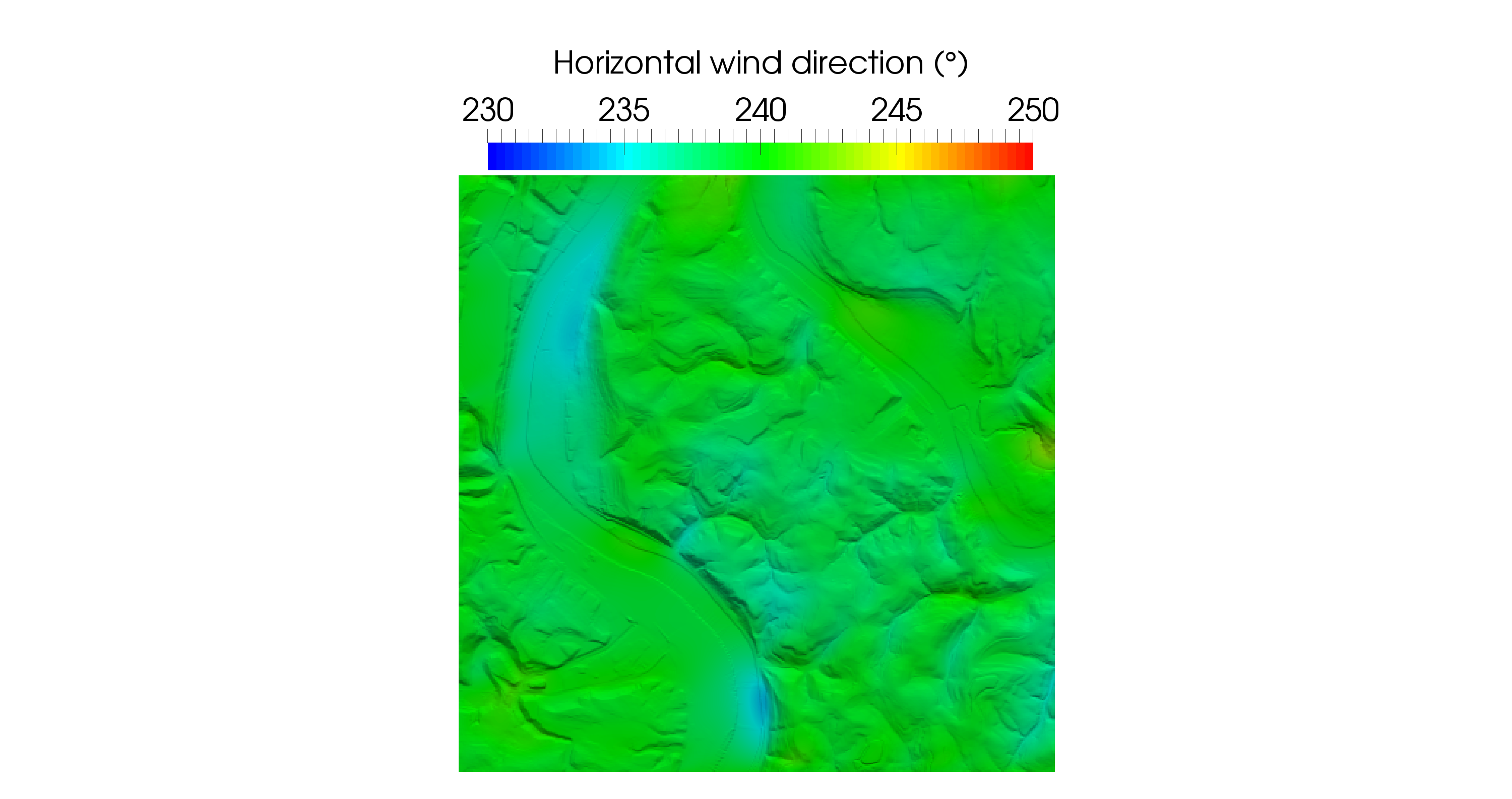}
    \caption{}
    \label{fig:contour_u_dir_200m}
  \end{subfigure}
  \caption{Contours of horizontal wind speed and wind direction at different heights AGL. (a) Wind speed ($m/s$), 10 $m$ AGL.(b) Wind direction (°), 10 $m$ AGL. (c) Wind speed ($m/s$), 200 $m$ AGL. (d) Wind direction (°), 200 $m$ AGL.}
  \label{fig:contour_u}
\end{figure}

The effects of the irregular topography on flow patterns are studied in terms of wind speed and wind direction. Figure \ref{fig:contours_z_vector} shows contours of the height above the ground and the wind speed vectors colored by the magnitude of the vertical wind speed. On the southwest side of the Monongahela River where the terrain is relatively flat, the vertical wind speed is near zero and wind direction is close to uniform. The wind vectors show that wind will either climb over the hills or enter the channels of the valleys when traveling from southwest to northeast. Once entering the channels, the wind tends to move along the channels until it meets the next hill. Figures \ref{fig:contour_u_10m} to \ref{fig:contour_u_dir_200m} show contours of horizontal wind speed and wind direction extracted at different heights AGL. At 10 $m$ AGL, wind speed and wind direction are strongly influenced by the terrain. High wind speed regions can be found on the windward-facing slopes of the hills, while low wind speed regions are found in valleys on the leeward side due to flow separation. These tendencies are in agreement with other studies of flow over 2D and 3D hills \citep{Sldek2007,Bechmann2010,Liu2016,Li2017,Berg2018}. The dominant wind direction is the same as the wind direction set at the inlet of the domain. The regions where the wind directions differ the most from the dominant wind direction are also the regions where sudden changes in height are present. At 200 $m$ AGL, wind speed and wind direction become more uniform and the effects of the topography are less significant. Without the irregular topography, the wind speed contours will be uniform at different height levels AGL, emphasizing the importance of the inclusion of the real-world topography in the simulations of wind development.

\begin{figure}[hbt!]
  \centering
  \begin{subfigure}{0.24\linewidth}
    \centering
    %%trim=left bottom right top  
    \includegraphics[trim=900 120 890 190,clip,width=1.0\linewidth]{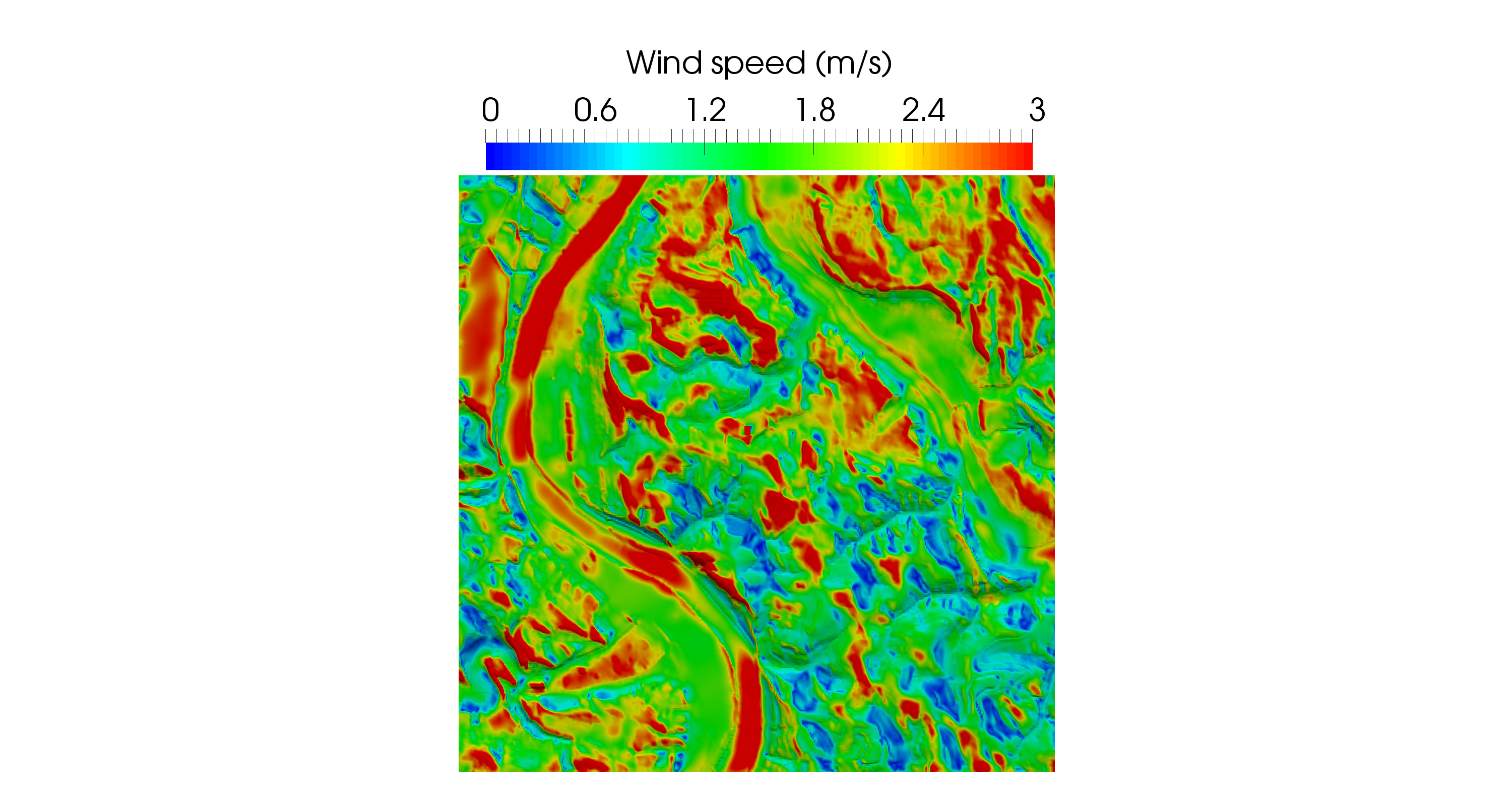}
    \caption{}
    \label{fig:contour_u_bottom_z0_variable}
  \end{subfigure}
  \begin{subfigure}{0.24\linewidth}
    \centering
    \includegraphics[trim=900 120 890 190,clip,width=1.0\linewidth]{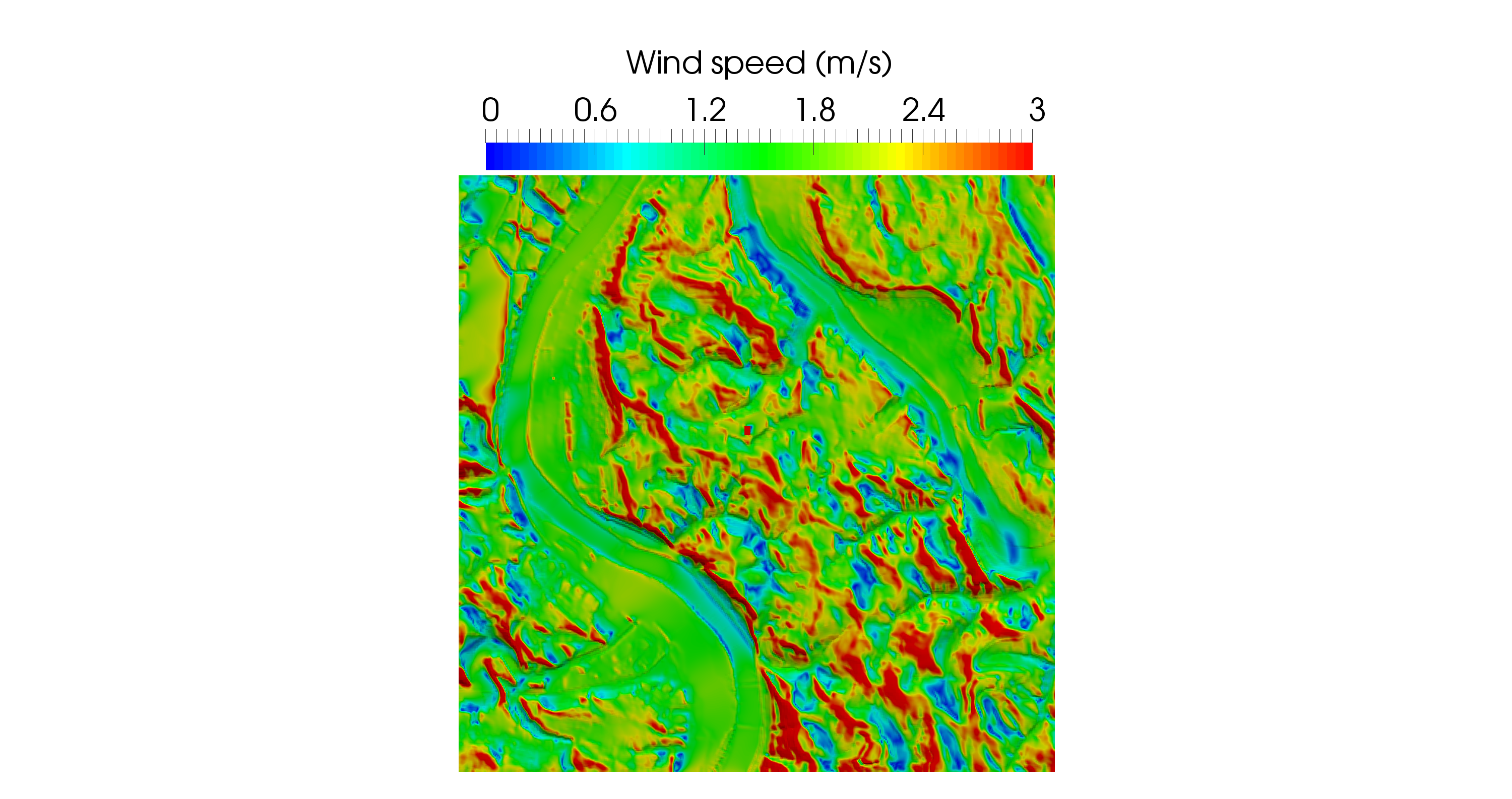}
    \caption{}
    \label{fig:contour_u_bottom_z0_uniform}
  \end{subfigure}
  \begin{subfigure}{0.24\linewidth}
    \centering
    %%trim=left bottom right top  
    \includegraphics[trim=900 120 890 190,clip,width=1.0\linewidth]{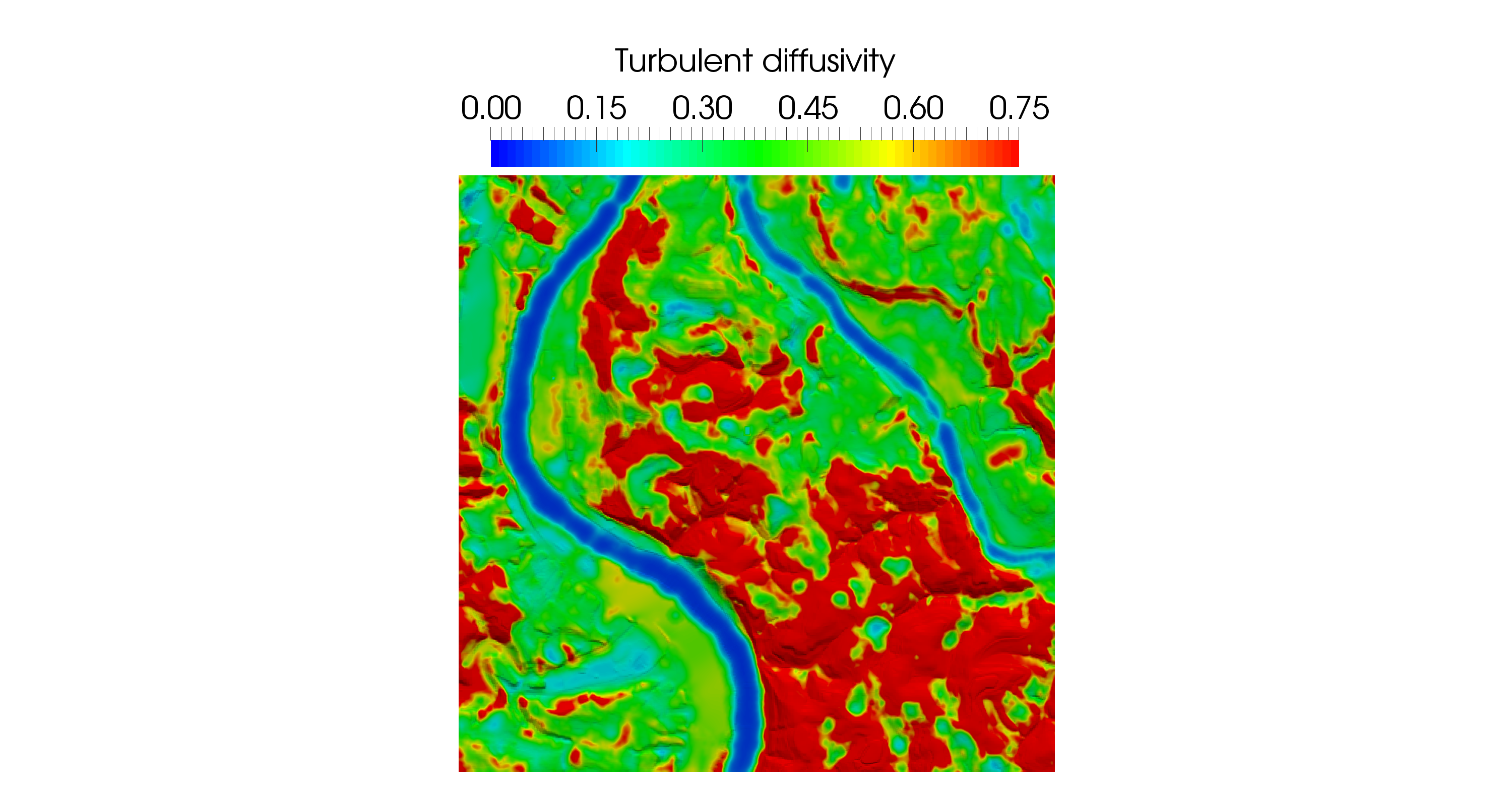}
    \caption{}
    \label{fig:contour_nut_bottom_z0_variable}
  \end{subfigure}
  \begin{subfigure}{0.24\linewidth}
    \centering
    \includegraphics[trim=900 120 890 190,clip,width=1.0\linewidth]{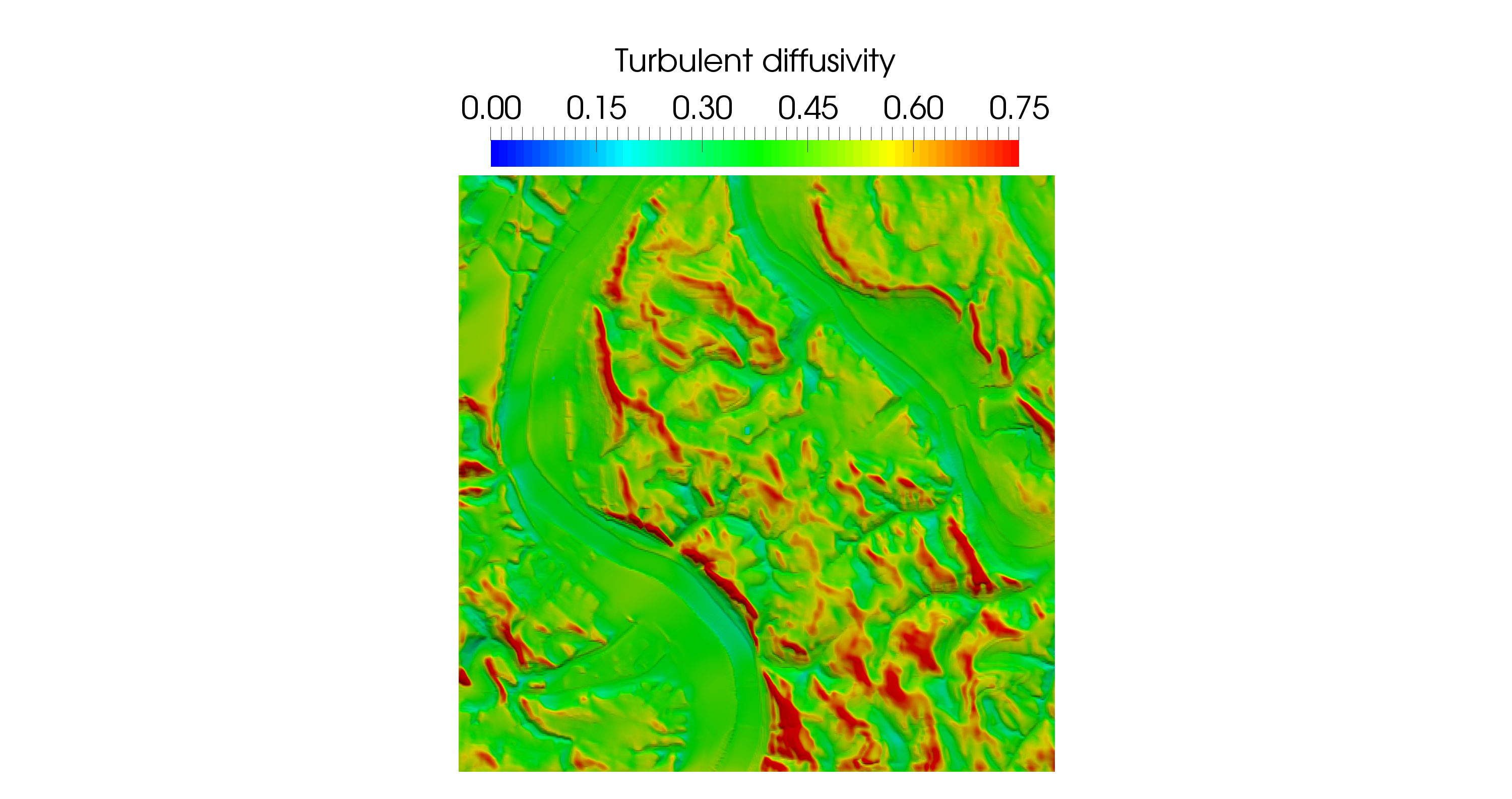}
    \caption{}
    \label{fig:contour_nut_bottom_z0_uniform}
  \end{subfigure}
  \caption{Contours of horizontal wind speed ($m/s$) using (a) variable and (b) uniform $z_0$ and contours of turbulent diffusivity ($m^2/s$)  using (c) variable and (d) uniform $z_0$ in the first cells AGL.}
  \label{fig:contour_var_uni}
\end{figure}

The current model implements variable surface roughness ($z_0$) in the entire computational domain. In order to compare the impact of variable $z_0$ on wind development and turbulence generation, we have also performed one simulation with uniform $z_0$. The value of uniform $z_0$ is taken to be the mean value of the variable $z_0$, which is around 0.7 $m$. Since $z_0$ is a parameter specified in the wall model, the differences between the results of the variable $z_0$ and that of the uniform $z_0$ are expected to be most significant near the ground. Figures \ref{fig:contour_u_bottom_z0_variable} and \ref{fig:contour_u_bottom_z0_uniform} show contours of horizontal wind speed and Figures \ref{fig:contour_nut_bottom_z0_variable} and \ref{fig:contour_nut_bottom_z0_uniform} show contours of turbulent diffusivity in the first layer of cells AGL for variable and uniform $z_0$. 

\begin{figure}[htb!]
  \centering
  \begin{subfigure}{0.5\linewidth}
    \centering
    %%trim=left bottom right top  
    \includegraphics[width=0.9\linewidth]{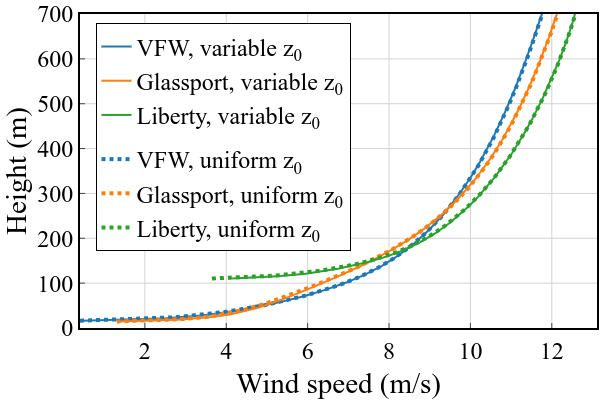}
    \caption{}
    \label{fig:wind}
  \end{subfigure}%
  \begin{subfigure}{0.5\linewidth}
    \centering
    \includegraphics[width=0.9\linewidth]{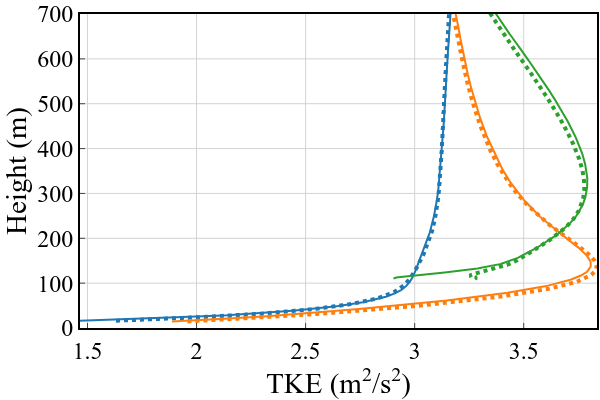}
    \caption{}
    \label{fig:k_nolegend}
  \end{subfigure}
  \begin{subfigure}{0.5\linewidth}
    \centering
    \includegraphics[width=0.9\linewidth]{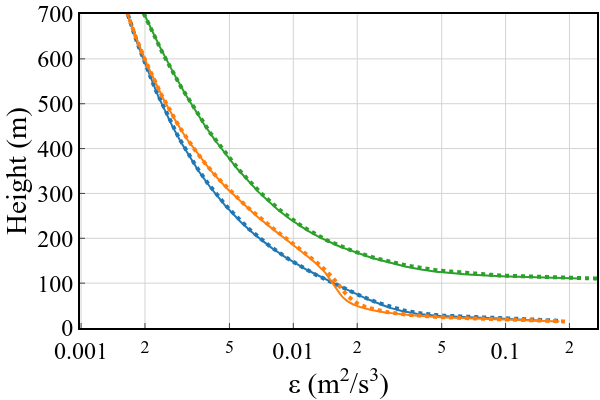}
    \caption{}
    \label{fig:epsilon_nolengend}
  \end{subfigure}%
  \begin{subfigure}{0.5\linewidth}
    \centering
    \includegraphics[width=0.9\linewidth]{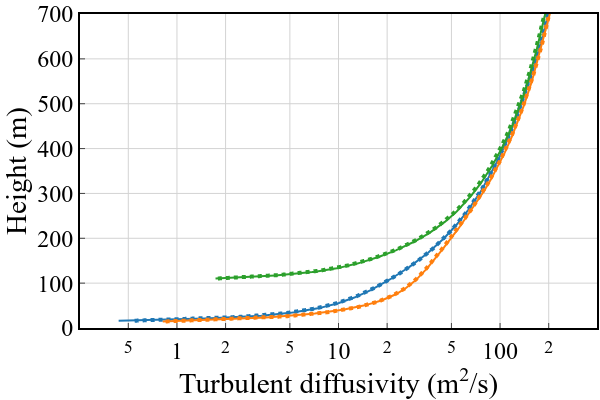}
    \caption{}
    \label{fig:nut_nolegend}
  \end{subfigure}
  \caption{Vertical profiles of (a) wind speed, (b) TKE, (c) the dissipation rate of TKE, and (d) turbulent diffusivity at different locations under variable and uniform $z_0$.}
  \label{fig:wind_tke_e_nut}
\end{figure}
From Figures \ref{fig:contour_u_bottom_z0_variable} and \ref{fig:contour_u_bottom_z0_uniform}, the difference in the horizontal wind speed is relatively small. The two rivers have near zero value of $z_0$. After using a uniform value of 0.7 $m$, the wind speeds on the rivers drop a lot. For turbulent diffusivity, Figures \ref{fig:contour_nut_bottom_z0_variable} and \ref{fig:contour_nut_bottom_z0_uniform} show that result from variable $z_0$ near the southeast corner of the domain where forests exist has larger values compared to that of uniform $z_0$. However, these differences are only obvious near the ground. As seen from vertical profiles of horizontal wind speed, turbulent kinetic energy (TKE) and its dissipation ($\varepsilon$), and the turbulent diffusivity in Figure \ref{fig:wind_tke_e_nut}, variable and uniform $z_0$ yield very similar wind and turbulence results at different locations. Therefore, it is appropriate to specify a reasonable value of uniform $z_0$ representing the local land-use patterns.
\subsection{Model predictions of \ch{SO2} dispersion}

\begin{figure}[htb!]
  \centering
  \begin{subfigure}{0.5\linewidth}
    \centering
    %%trim=left bottom right top  
    \includegraphics[width=0.9\linewidth]{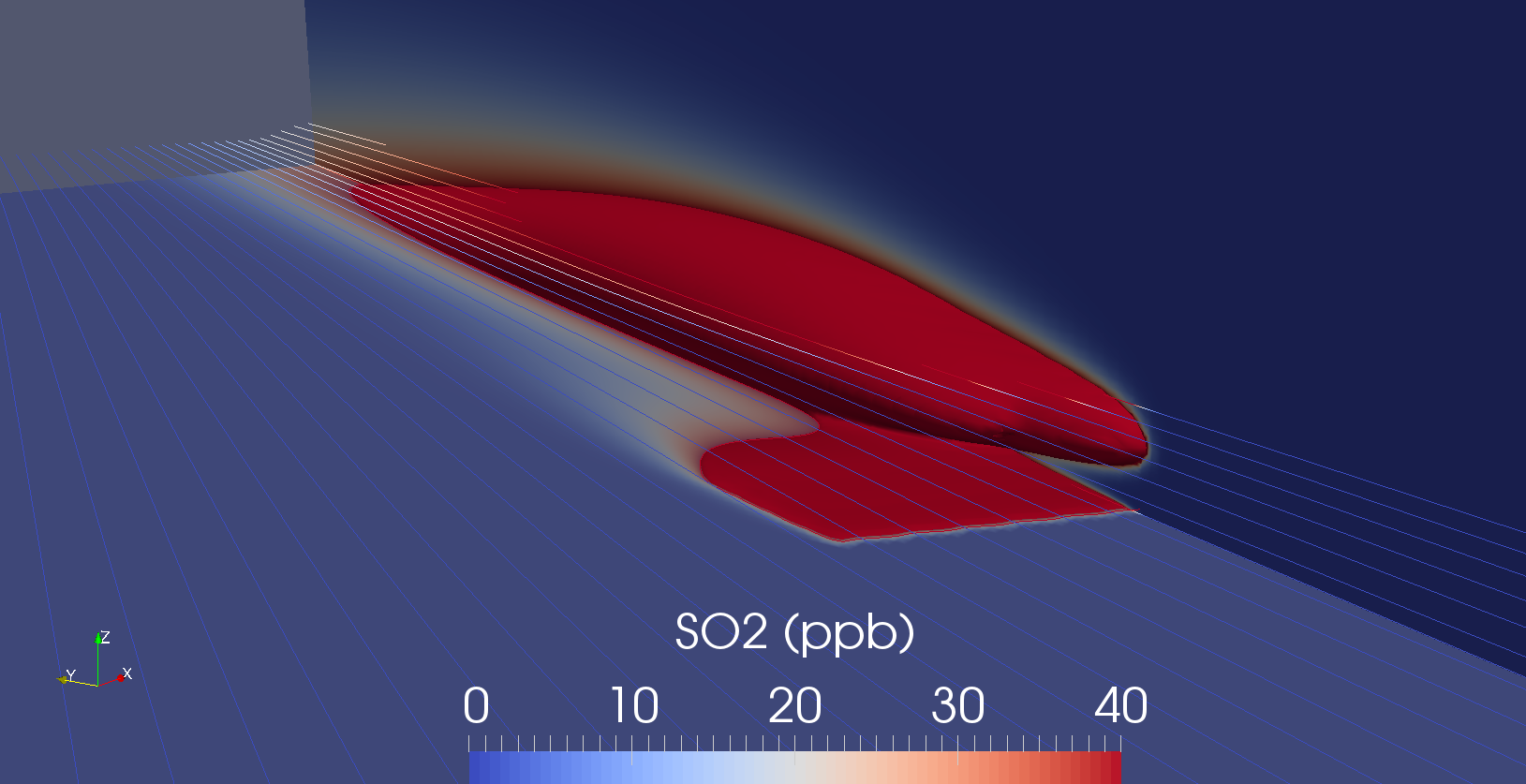}
    \caption{}
    \label{fig:neutral_flat}
  \end{subfigure}%
  \begin{subfigure}{0.5\linewidth}
    \centering
    \includegraphics[width=0.9\linewidth]{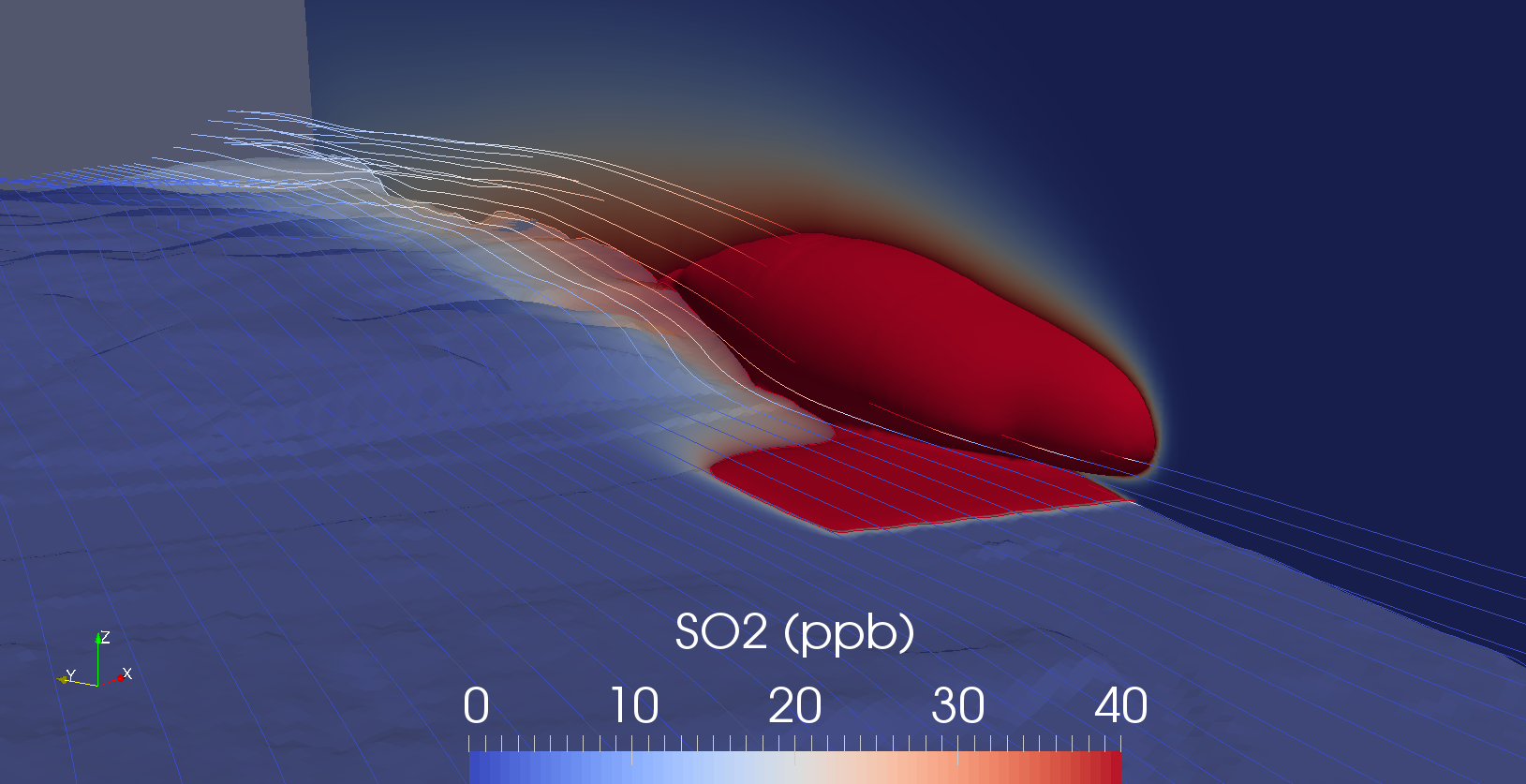}
    \caption{}
    \label{fig:neutral_terrain}
  \end{subfigure}
  \begin{subfigure}{0.5\linewidth}
    \centering
    %%trim=left bottom right top  
    \includegraphics[width=0.9\linewidth]{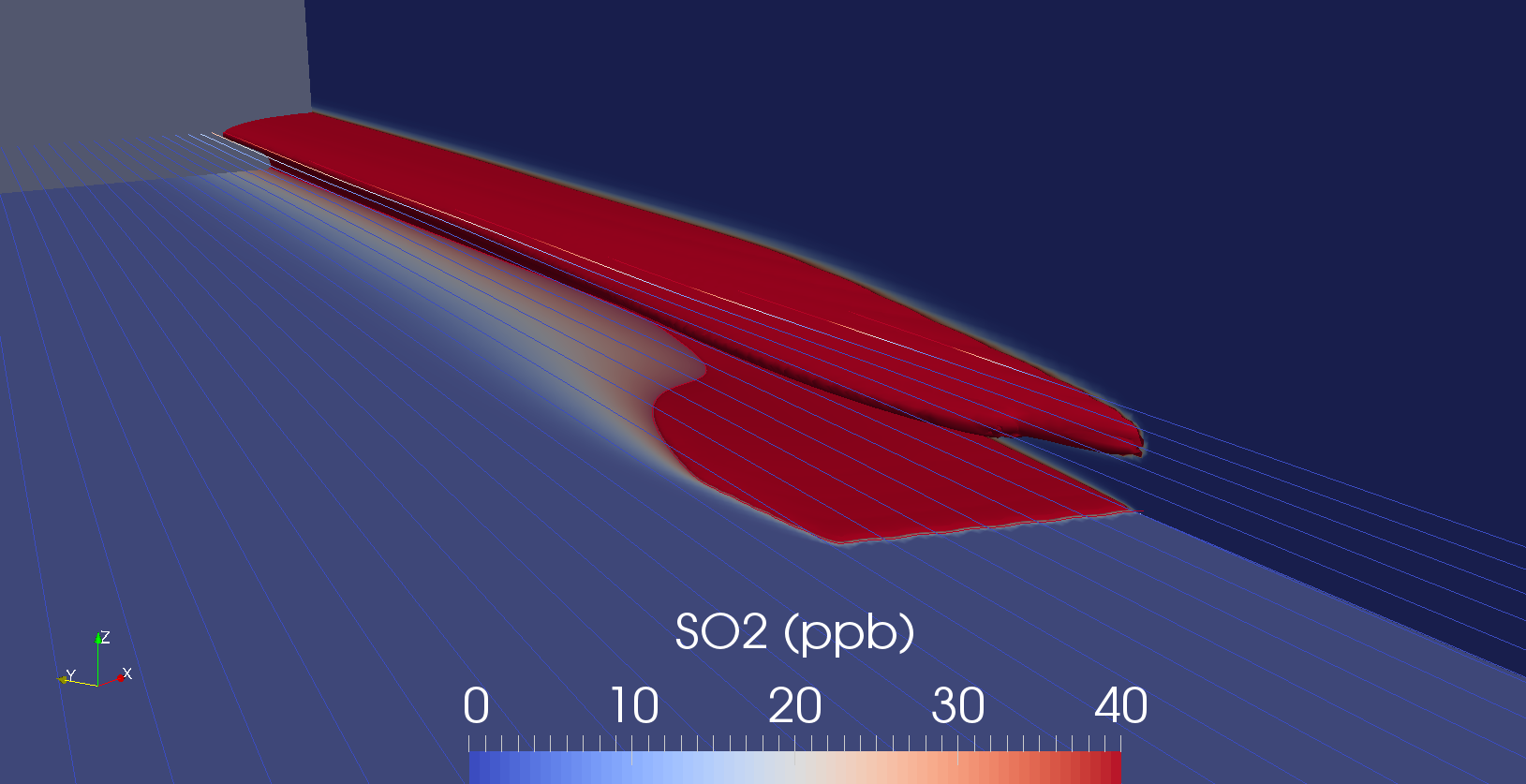}
    \caption{}
    \label{fig:stable_flat}
  \end{subfigure}%
  \begin{subfigure}{0.5\linewidth}
    \centering
    \includegraphics[width=0.9\linewidth]{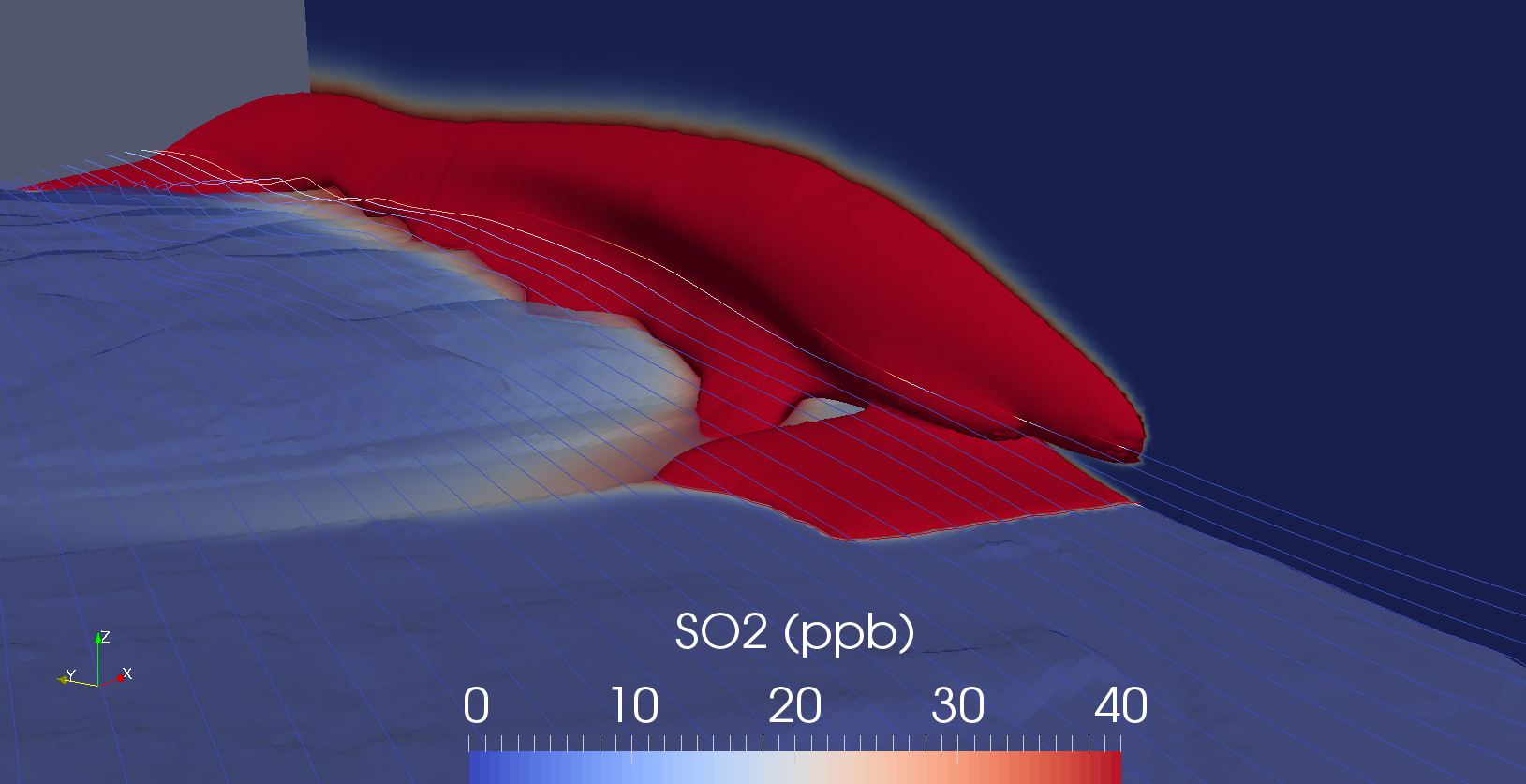}
    \caption{}
    \label{fig:stable_terrain}
  \end{subfigure}
  \caption{CFD results under different thermal conditions over flat and complex terrain. (a) Flat terrain under neutral condition. (b) Complex terrain under neutral condition. (c) Flat terrain under stable condition. (d) Complex terrain under stable condition. Contours show \ch{SO2} concentrations on the ground, isosurfaces created from 40 ppb, and slice cut passing through the center of the plant and the Liberty monitor. Streamtraces are also shown.}
  \label{fig:neutral_stable_flat_terrain}
\end{figure}

The current National Ambient Air Quality Standards for \ch{SO2} is 75 parts per billion (ppb) based on the 3-year average of the 99th percentile of the yearly distribution of 1-hour daily maximum concentrations \cite{EPA_NAAQS}. In the present work, a steady-state CFD model is used to represent hourly periods of emissions assuming that the boundary conditions do not change over the selected 1-hour periods. About 90\% of the \ch{SO2} is emitted from the ten stacks and remaining at the ground level. After being released from the plant, \ch{SO2} travels in the form of a plume that widens with distance from the plant. 
The complexity of the real-world terrain plays an important role in the predictions of the \ch{SO2} plume structure. Figure \ref{fig:neutral_stable_flat_terrain} shows the development of the plume under neutral and stable conditions, with the actual complex terrain and if the terrain were simplified to a hypothetical flat surface. In case of the simplified flat terrain, the plume centerline stays at around the same height of the stack exit. For the actual terrain, the plume goes through hills and valleys giving rise to a more complex structure. The differences are larger for the stable condition compared to the neutral condition. These results suggest that specifying the complexity of the terrain in the model is important for predictions of \ch{SO2} dispersion.

\begin{figure}[htb!]
  \centering
\begin{subfigure}{0.5\linewidth}
    \centering
    \includegraphics[width=0.9\linewidth]{fig/fig_neutral_terrain.png}
    \caption{}
    \label{fig:neutral_terrain_noUT}
  \end{subfigure}%
  \begin{subfigure}{0.5\linewidth}
    \centering
    %%trim=left bottom right top  
    \includegraphics[width=0.9\linewidth]{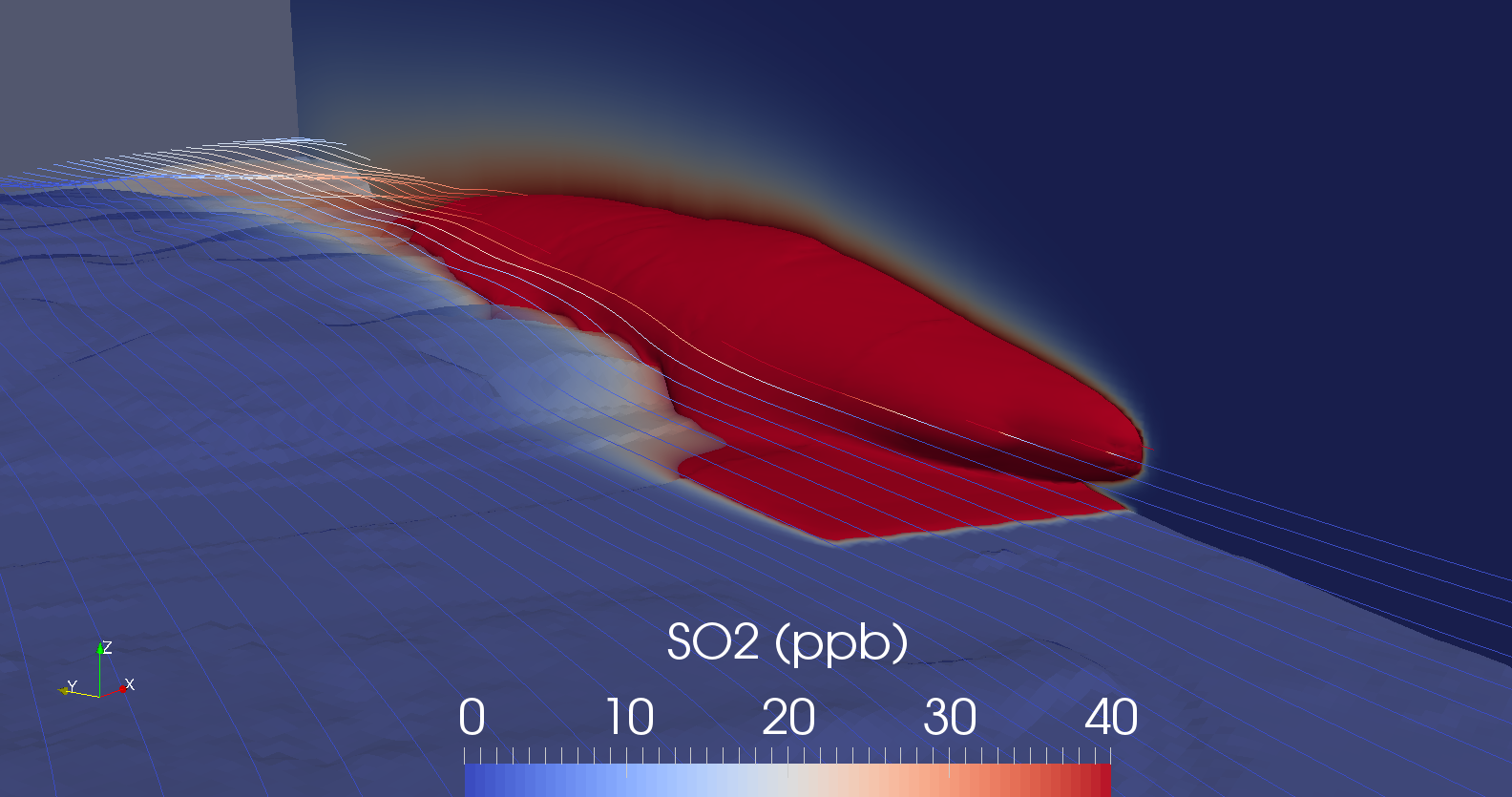}
    \caption{}
    \label{fig:neutral_bSF_noUT}
  \end{subfigure}
  \begin{subfigure}{0.5\linewidth}
    \centering
    %%trim=left bottom right top  
    \includegraphics[width=0.9\linewidth]{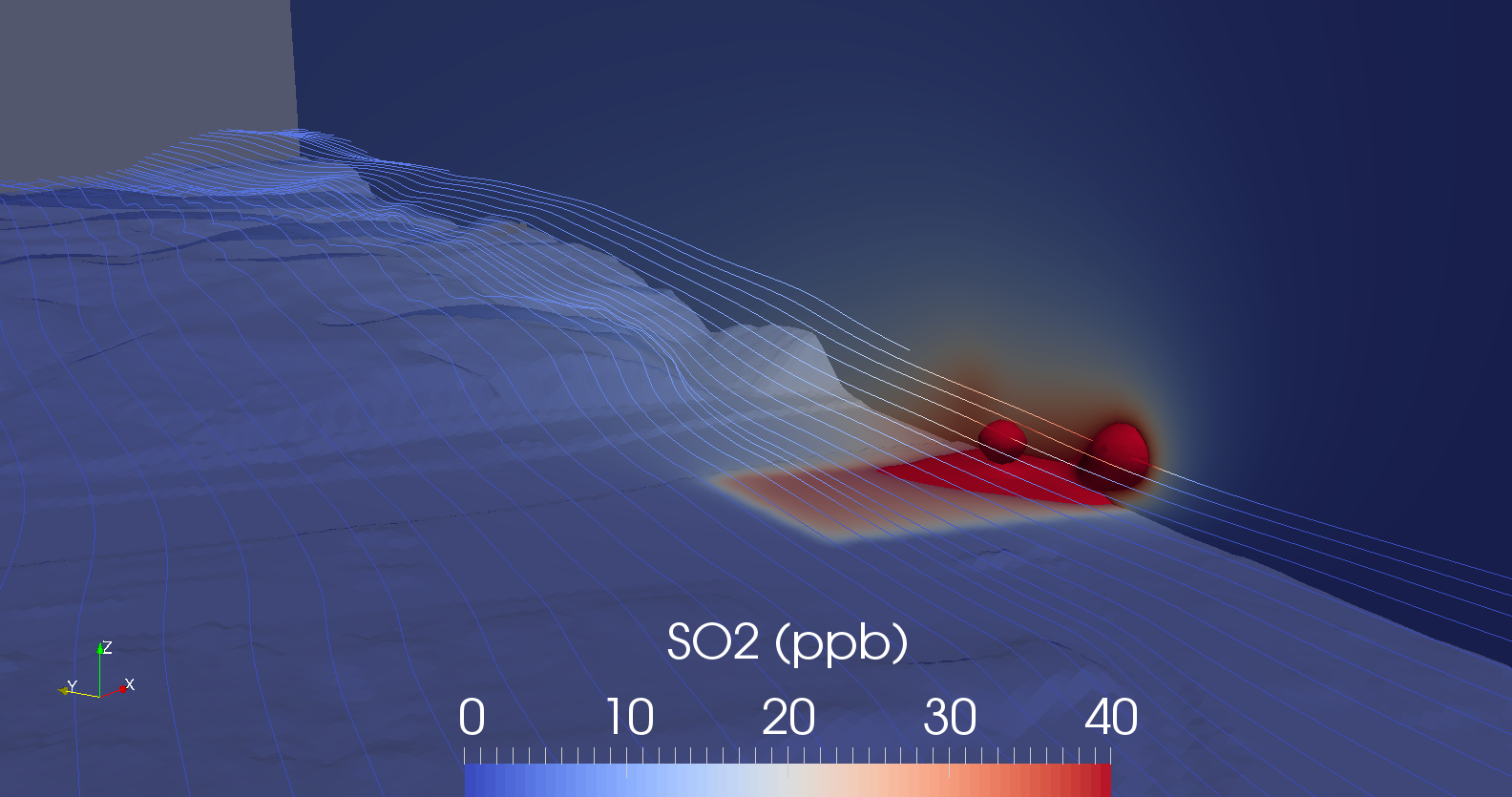}
    \caption{}
    \label{fig:neutral_bBSF_UT}
  \end{subfigure}%
  \begin{subfigure}{0.5\linewidth}
    \centering
    \includegraphics[width=0.9\linewidth]{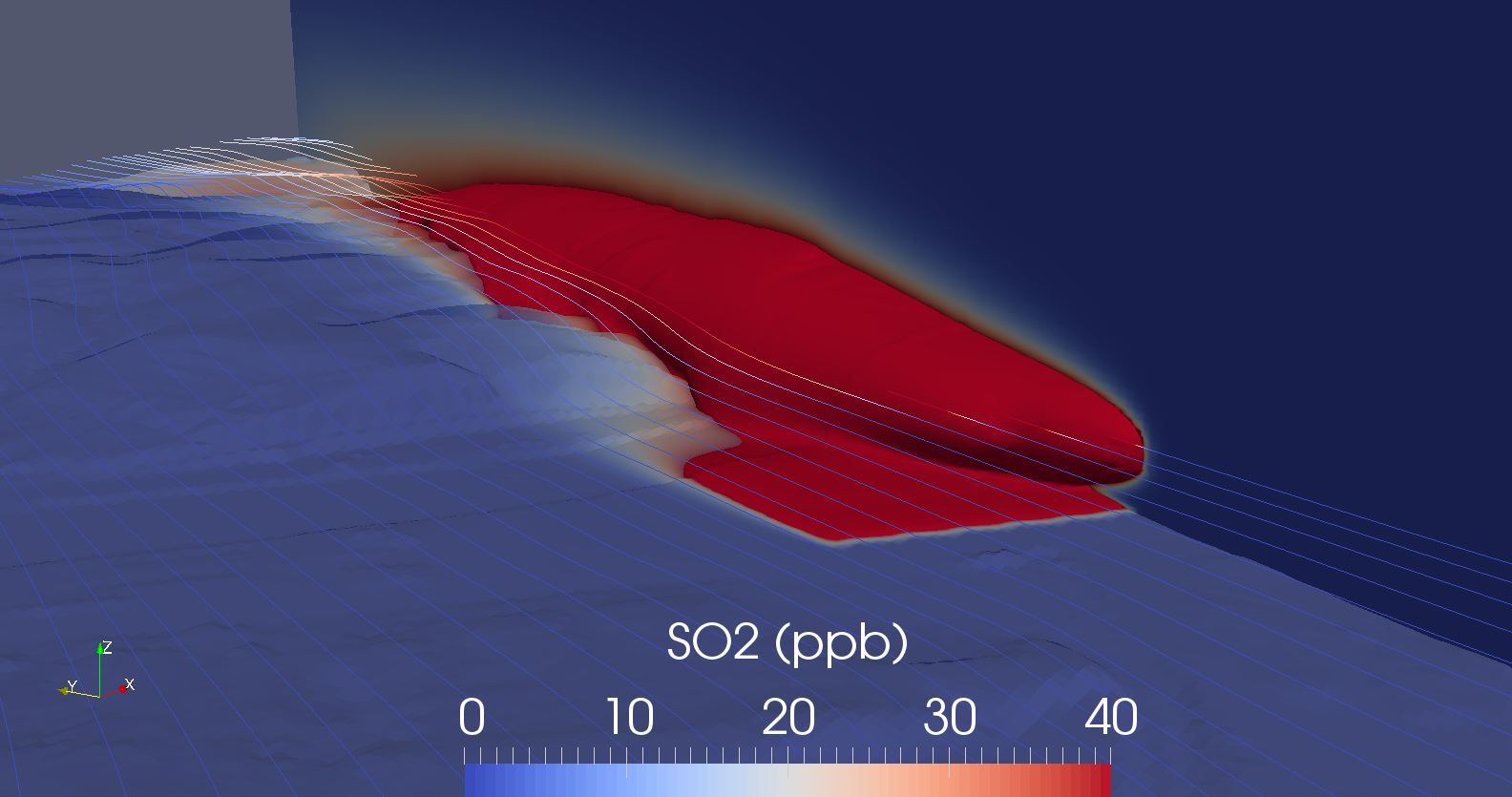}
    \caption{}
    \label{fig:neutral_bBSF_noUT}
  \end{subfigure}
  \caption{CFD results from different solvers with different stack exit conditions under neutral stability. (a) Compressible solver and setting the annualized stack temperatures (around $500K$) at the stack exits. (b) Compressible solver and setting the ambient temperature at the stack exits. (c) Incompressible solver with Boussinesq approximation and setting the annualized stack temperatures (around $500K$) at the stack exits. (d) Incompressible solver with Boussinesq approximation and setting the ambient temperature at the stack exits. Contours show \ch{SO2} concentrations on the ground, isosurfaces created from 40 ppb, and slice cut passing through the center of the plant and the Liberty monitor. Streamtraces are also shown.}
  \label{fig:neutral_bSF_bBSF_UT}
\end{figure}

\begin{figure}[htb!]
  \centering
  \begin{subfigure}{0.5\linewidth}
    \centering
    %%trim=left bottom right top  
    \includegraphics[width=0.9\linewidth]{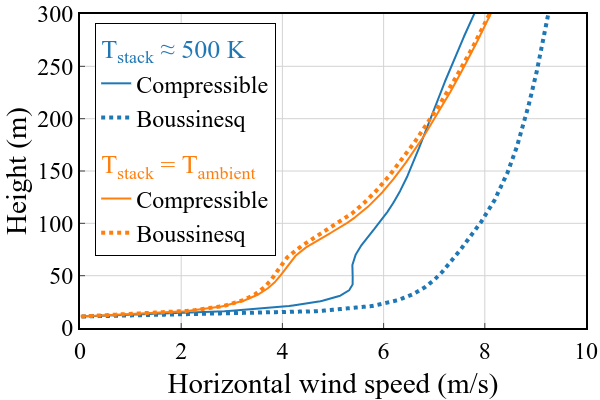}
    \caption{}
    \label{fig:bSF_bBSF_Uxy_UT_noUT_plant}
  \end{subfigure}%
  \begin{subfigure}{0.5\linewidth}
    \centering
    \includegraphics[width=0.9\linewidth]{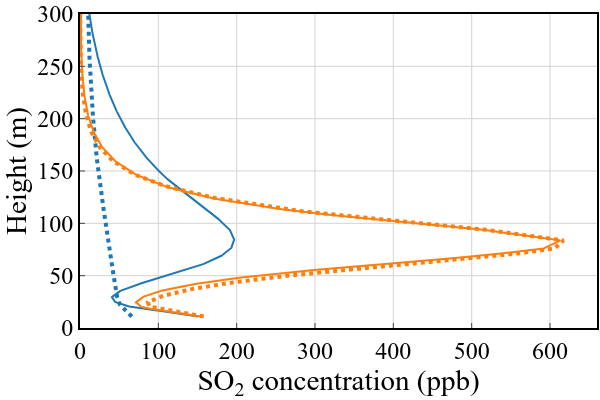}
    \caption{}
    \label{fig:bSF_bBSF_SO2_UT_noUT_plant_nolegend}
  \end{subfigure}
  \caption{Vertical profiles of (a) horizontal wind speed and (b) \ch{SO2} concentration at the plant center using different solvers with different stack exit conditions under neutral stability.}
  \label{fig:bSF_bBSF_Uxy_SO2_UT_noUT_plant}
\end{figure}

In addition to the complex terrain, it is also important to specify correct temperature boundary conditions at the exit of the stacks. Measurements suggest that the flue gas exits the stacks at an average temperature of about 500 $K$. Since the mean temperature in the domain ranges between 273 $K$ and 296 $K$ from different months \citep{NWS_mean} , there is a significant gradient of temperature between the atmosphere and the stack exit. Such temperature gradient will give rise to buoyancy inside the plume, especially near the stack exits. As discussed previously, buoyancy can be simulated using the full compressible flow equations or by using the Boussinesq approximation with the incompressible flow solver. In order to investigate the sensitivity of the predictions to model assumptions and stack exit conditions, simulations are performed with the compressible and incompressible flow solvers using two different stack exit temperatures. Figure \ref{fig:neutral_bSF_bBSF_UT} shows the predicted plumes under various model assumptions and stack exit conditions. Figure \ref{fig:bSF_bBSF_Uxy_SO2_UT_noUT_plant} shows the vertical profiles of the horizontal wind speed and \ch{SO2} extracted at the center of the coke plant. 

It can be seen that the difference between the compressible solver and the Boussinesq approximation is small when the stack exit temperature is set to ambient temperature. However, the differences between the two solvers become significant when stack exit temperatures are set to their measured values, suggesting that the compressible flow solver should be used when simulating the dispersion of pollutants emitted at elevated temperatures.  

To evaluate the performance the CFD model, the model predictions of \ch{SO2} dispersion need to be compared with available measurements from an \ch{SO2} monitor installed at the Liberty location shown in Figure \ref{fig:domain_sensor_pole}. The steady-state CFD model has provided reasonably good agreement for wind speed as discussed above. Over the selected time periods, the wind speed is relatively steady but the wind direction can change. Since we are comparing model predictions with a single-point measurement in entire the study domain, it is important to account for the uncertainty in the wind direction . A $5^{\circ}$ change in wind direction leads to about 200 $m$ of plume center displacement at the Liberty monitor, which is around 2 $km$ away from the center of the plant. To account for the uncertainty in the wind direction, \ch{SO2} concentration in the CFD model is sampled over an arc passing through the Liberty monitor. The sampling arc has been used by other researchers to compare model predictions with measurements \citep{Demael2008,Rood2014}. The sampling arc created in this work has a length of 3 $km$ with the Liberty monitor in the middle of the arc. The radius of the arc is about 2 $km$, which is the straight line distance from the center of the plant to the Liberty site. 

\begin{figure}[htb!]
  \centering
  \begin{subfigure}{0.5\linewidth}
    \centering
    %%trim=left bottom right top  
    \includegraphics[width=0.9\linewidth]{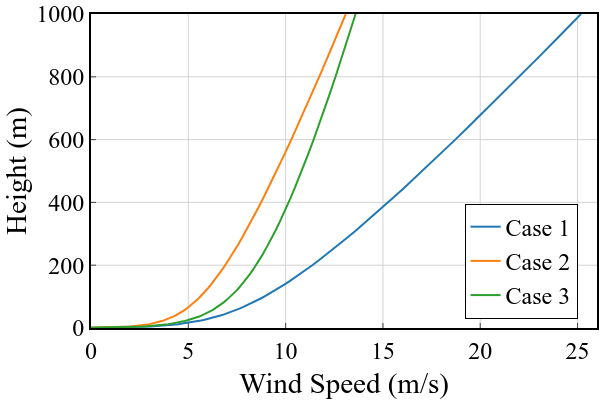}
    \caption{}
    \label{fig:3cases_speed}
  \end{subfigure}%
  \begin{subfigure}{0.5\linewidth}
    \centering
    \includegraphics[width=0.9\linewidth]{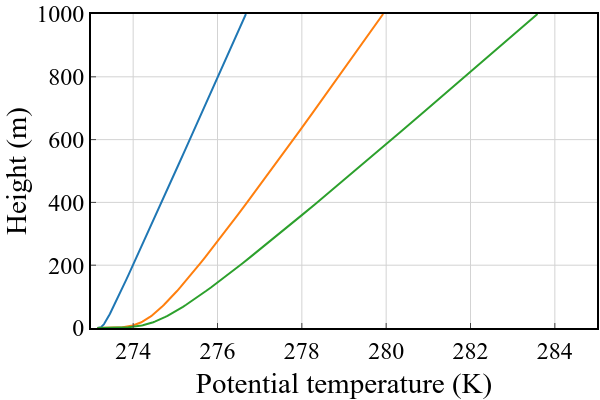}
    \caption{}
    \label{fig:3cases_temperature_nolegend}
  \end{subfigure}
  \caption{Fitted inlet profiles of (a) wind speed and (b) potential temperature for the three cases.}
  \label{fig:3cases_speed_temperature}
\end{figure}

Predictions of the model are compared with the measurements for three cases that represent different wind speeds and stability classes. Vertical profiles of horizontal wind speed and potential temperature for the three cases are shown in Figure \ref{fig:3cases_speed_temperature}. The wind directions reported by the Liberty monitor for the three cases are within $6^{\circ}$ from each other as shown in Table \ref{tab:5}.  To better compare the strength of the inversion among the three cases, the ground temperatures are adjusted so that they all start at 273.15 $K$. In the actual simulations, each case uses it own ground temperature specified using measurements from the KAGC location as shown in Figure \ref{fig:domain_wind_data}. For the three cases, case 1 has higher wind speed compared to the other two cases. Case 1 has the weakest inversion, case 2 has moderate inversion, while case 3 has the strongest inversion as suggested by the slopes of the potential temperatures. The emission rate for each case is shown in Table \ref{tab:5}. 

\begin{table}[hbt!]
  \centering
  \captionof{table}{Summary of the three cases with different wind conditions and inversion strengths.}
  \scalebox{0.9}{
\begin{tabular}{ccccccc} 
\hline
Case & \begin{tabular}[c]{@{}c@{}}Wind\\direction \\($^{\circ}$)\end{tabular} & \begin{tabular}[c]{@{}c@{}}Emission \\rate \\(g/s)\end{tabular} & \begin{tabular}[c]{@{}c@{}}Monitor: 1-hour \\average data \\(ppb)\end{tabular} & \begin{tabular}[c]{@{}c@{}}CFD: extracted \\value \\at Liberty \\(ppb)\end{tabular} & \begin{tabular}[c]{@{}c@{}}Monitor: max \\1-minute \\average data\\(ppb)\end{tabular} & \begin{tabular}[c]{@{}c@{}}CFD:~max value \\on arc \\(ppb)\end{tabular}  \\ 
\hline
1    & 209                                                                    & 234                                                             & 43                                                                             & 53                                                                                  & 58                                                                                    & 54                                                                       \\
2    & 208                                                                    & 277                                                             & 39                                                                             & 38                                                                                  & 61                                                                                    & 40                                                                       \\
3    & 214                                                                    & 209                                                             & 43                                                                             & 46                                                                                  & 54                                                                                    & 78                                                                       \\
\hline
\end{tabular}}
  \label{tab:5}
\end{table}

\begin{figure}[htb!]
  \centering
  \begin{subfigure}{0.5\linewidth}
    \centering
    %%trim=left bottom right top  
    \includegraphics[width=0.9\linewidth]{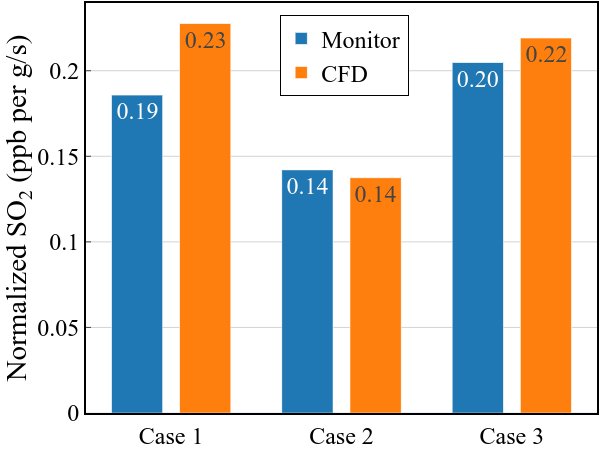}
    \caption{}
    \label{fig:3cases_barChart_normalized_Liberty_SO2}
  \end{subfigure}%
  \begin{subfigure}{0.5\linewidth}
    \centering
    \includegraphics[width=0.9\linewidth]{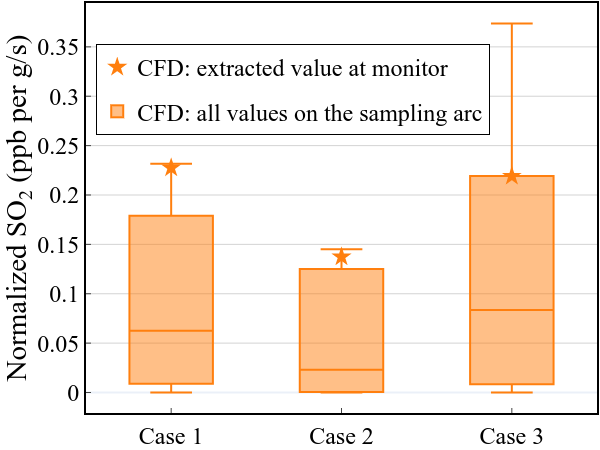}
    \caption{}
    \label{fig:3cases_box_normalized_CFD_SO2}
  \end{subfigure}
  \caption{(a) Bar chart of \ch{SO2} concentrations from the 1-hour average Liberty measurements and the steady-state CFD predictions extracted at Liberty. (b) Box plot of all the values on the sampling arc and the extracted value (marked by $\bigstar$) at Liberty. The concentrations (ppb) are normalized by the corresponding emission rates (g/s).}
  \label{fig:3cases_bar_normalized}
\end{figure}
In Table \ref{tab:5}, both measured and predicted concentrations are presented as absolute ppb. In addition, Figure \ref{fig:3cases_bar_normalized} shows the concentrations from CFD predictions and measurement that are normalized by their respective emission rates.  As can be seen from Table \ref{tab:5}, the 1-hour average \ch{SO2} concentrations agree well with the CFD model predictions extracted at the exact Liberty monitor location. From Figure \ref{fig:3cases_barChart_normalized_Liberty_SO2}, after normalization, case 2 has the lowest concentration from both the monitor and the CFD model. Using the sampling arc, we want to locate the highest concentration near the Liberty monitor. The maximum values on the sampling arc are compared with the values extracted at the Liberty monitor as shown in Figure \ref{fig:3cases_barChart_normalized_Liberty_SO2}. The maximum concentrations on the sampling arc show the same trend as that of the extracted value at Liberty, with maximum concentration being higher for cases 1 and 3 and lowest for case 2. For cases 1 and 2, the maximum concentrations from the arc are very close to those of the extracted values at Liberty. However, in case 3, the maximum value of the arc is much larger, which indicates that the single monitor is not enough to capture the highest concentrations in the vicinity. 
 
\begin{figure}[hbt!]
  \centering
  \begin{subfigure}{0.5\linewidth}
    \centering
    %%trim=left bottom right top  
    \includegraphics[width=0.95\linewidth]{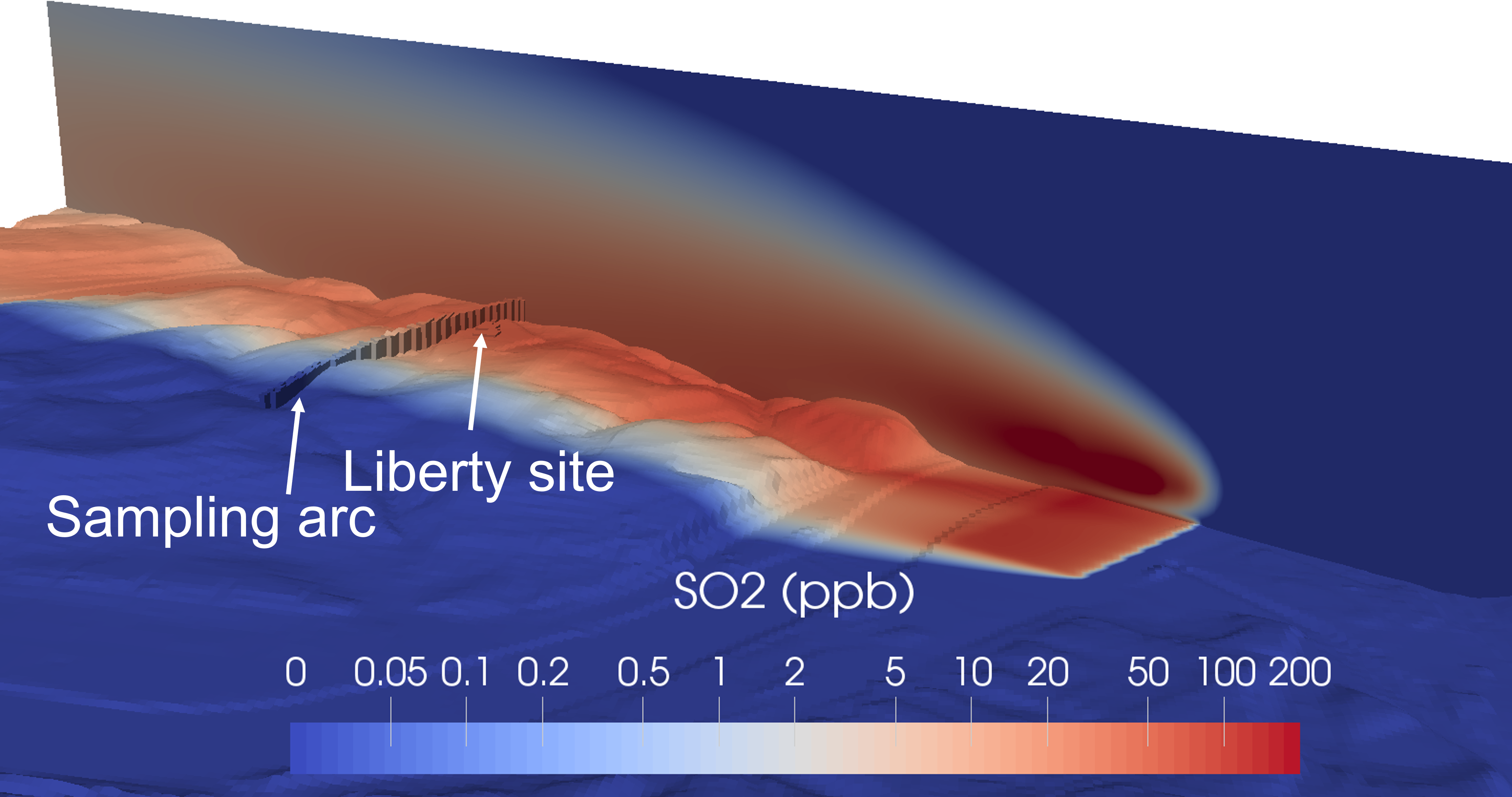}
    \caption{Case 1.}
    \label{fig:3cases_SO2_1}
  \end{subfigure}%
  \begin{subfigure}{0.5\linewidth}
    \centering
    \includegraphics[width=0.95\linewidth]{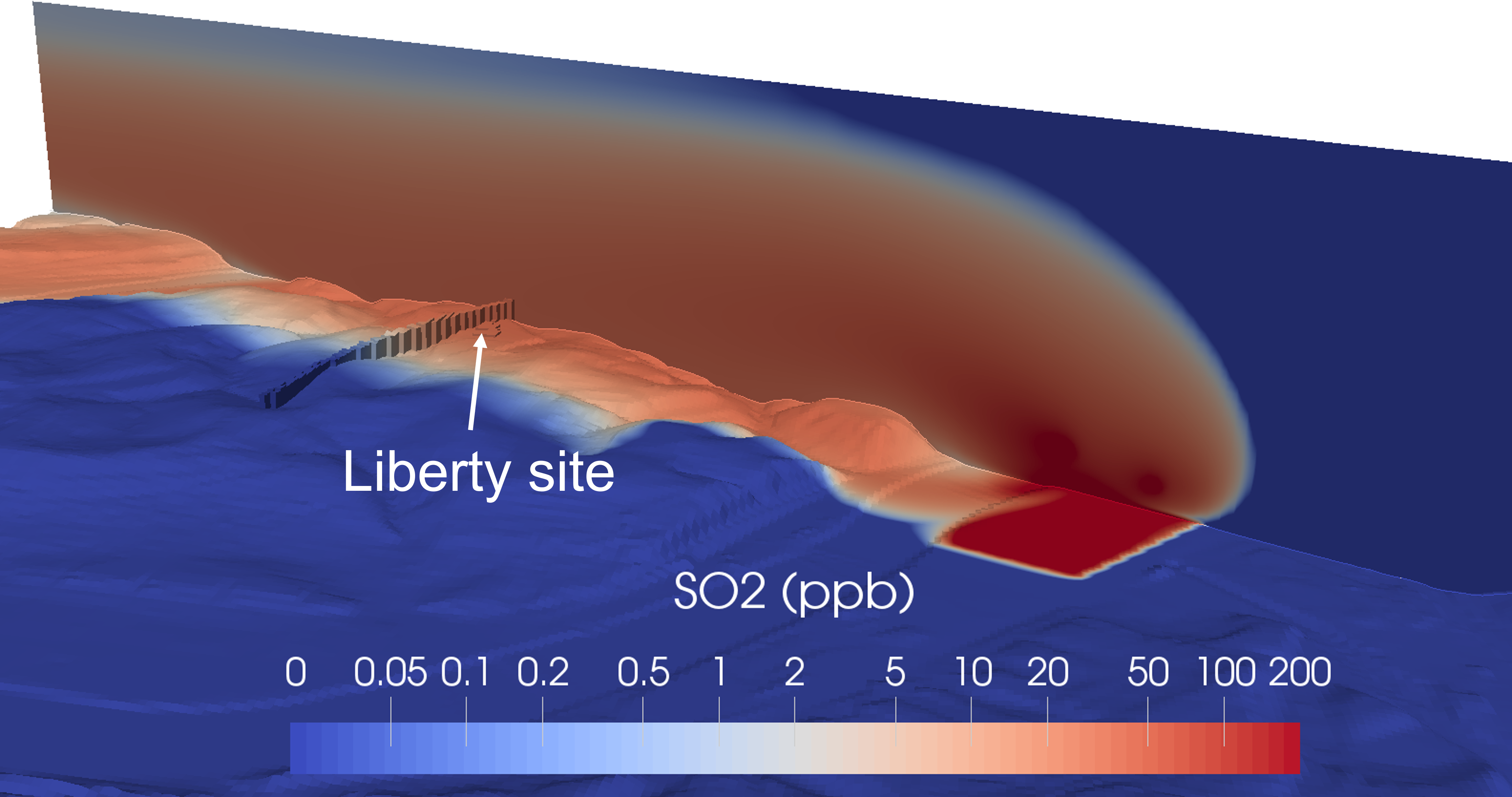}
    \caption{Case 2.}
    \label{fig:3cases_SO2_2}
  \end{subfigure}
  \begin{subfigure}{0.5\linewidth}
    \centering
    \includegraphics[width=0.95\linewidth]{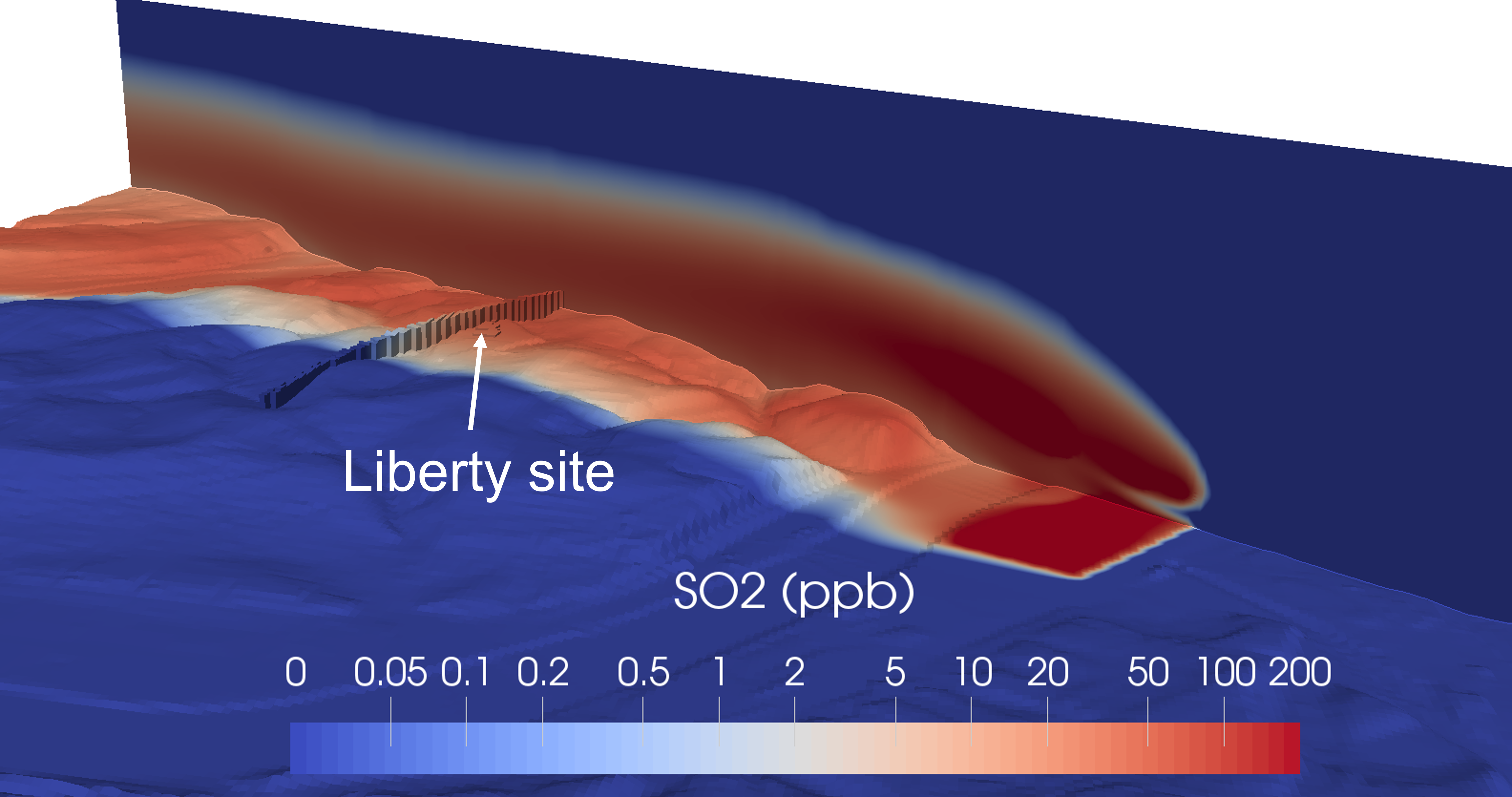}
    \caption{Case 3.}
    \label{fig:3cases_SO2_3}
  \end{subfigure}
  \caption{Contours of \ch{SO2} concentration (in log scale) on the ground and on the slice that passes through the plant center and aligns with the inlet wind direction.}
  \label{fig:3cases_SO2}
\end{figure}

The CFD model can be utilized to understand the factors leading to variable \ch{SO2} concentration at the measurement location. Case 1 has weaker inversion than case 2, but higher normalized \ch{SO2} concentration. Generally speaking, stronger inversion leads to higher pollution concentration with everything else being the same. However, for cases 1 and 2 presented here, the wind speed is significantly larger for case 1. Faster wind speed acts to bend the plume, thus \ch{SO2} is contained closer to the ground compared to case 1. The contours of \ch{SO2} concentration on a vertical 2D plane passing through the plume and on the ground surface are shown in Figure \ref{fig:3cases_SO2}. From Figure \ref{fig:3cases_SO2_1}, the plume does not disperse vertically, as would be expected for weaker inversion.  Vertical dispersion of the plume is more significant for case 2 as shown in Figure \ref{fig:3cases_SO2_2}, even though it has relatively stronger inversion. Case 3 has a slightly higher wind speed and stronger inversion compared to case 2. Both factors lead to higher emission normalized concentrations for case 3. As shown in Figure \ref{fig:3cases_SO2_3}, case 3 shows significantly lower vertical dispersion of the plume compared to case 2. Among the three cases, case 3 has the lowest vertical dispersion, leading to the highest concentrations inside the plume. These higher concentrations for case 3 are maintained for longer distances from the plant.  

\section{Conclusions}
In this study, wind development and \ch{SO2} dispersion within the ABL over a complex terrain is investigated through comparisons between CFD model predictions and field measurements. A CFD modeling methodology that includes the generation of a good quality fully-hex computational mesh, construction of boundary conditions using existing meteorological data, turbulence model parameterizations, and treatment of aerodynamic roughness length is discussed under different stability conditions.

The proposed curve-fitting method provides a consistent estimate of the atmosphere guided by the ground sensors and vertical profiles from sounding and reanalysis data. It can generate good quality inlet boundary conditions in terms of vertical profiles of horizontal wind speed and temperature for use in the CFD model. Validation of the CFD model against wind measurements suggests that the assumption of quasi steady-state periods seems reasonable. The CFD model agrees reasonably well with wind measurements for a number of different cases. It should, however, be mentioned that these cases were selected during periods of relatively small variations in wind speed and direction. If changes in wind speed and direction are more significant, the steady-state assumption will break down and a transient model will be needed.

The influence of the complex terrain in terms of topography on wind development can easily go up to 200 $m$ AGL. Near the ground, high wind speed regions can be found on the windward-facing slopes of the hills, while low wind speed regions are usually located in valleys on the leeward side. The topography of the complex terrain also generates more turbulence compared to that of a flat terrain. As for aerodynamic roughness length, its main influence is on the horizontal wind speed and turbulence near ground level. Above 10 $m$ AGL, using the variable and uniform aerodynamic roughness length gives almost the same wind and turbulence fields.

For \ch{SO2} dispersion, the compressible flow solver should be used due to the significant temperature gradient between the ambient environment and the stack exit. The developed CFD model is capable of distinguishing ABL flows of different inversion strengths and make reasonable predictions of the \ch{SO2} concentration when compared with data from the Liberty monitor. The model suggests that inversion is important for vertical dispersion of \ch{SO2}, and stronger inversion is associated with higher \ch{SO2} concentration. In addition to inversion, it is found that horizontal wind speed can lead to high concentrations at the monitor. The sampling arc is used to account for the uncertainty in wind direction, and it provides a more complete view of the \ch{SO2} concentration near the single monitor. When the monitor does not show high levels of \ch{SO2} concentrations, the sampling arc can help to check if high concentrations exist in the vicinity.

\section{Acknowledgments}
This study is made possible through sponsorship from the Allegheny County Health Department (ACHD). Special thanks to Tony Sadar and Jason Maranche of ACHD. The authors would also like to acknowledge the support from the Extreme Science and Engineering Discovery Environment (XSEDE) under grant number ASC110028 through the Pittsburgh Supercomputing Center (PSC).

\nolinenumbers
%\lipsum[3-65]
\bibliography{manuscript}
\end{document}